\PassOptionsToPackage{bookmarks=false}{hyperref}
\PassOptionsToPackage{table,xcdraw}{xcolor}
\documentclass[sigconf]{acmart}
\usepackage[table,xcdraw]{xcolor}
\usepackage{amsmath,amssymb,amsfonts}
\usepackage{algorithm}
\usepackage[noend]{algpseudocode}
\usepackage{graphicx}
\usepackage{textcomp}
\usepackage{tabularx}
\usepackage{booktabs}
\usepackage{enumitem}
\usepackage{url}
\usepackage{todonotes}
\usepackage{framed}
\usepackage{multirow}
\usepackage{pifont}
\usepackage{lipsum}
\usepackage{etoolbox}
\usepackage{listings}
\usepackage{float}

\newcommand{\squishlist}{
 \begin{list}{$\bullet$}
  { \setlength{\itemsep}{0pt}
     \setlength{\parsep}{3pt}
     \setlength{\topsep}{3pt}
     \setlength{\partopsep}{0pt}
     \setlength{\leftmargin}{1.5em}
     \setlength{\labelwidth}{1em}
     \setlength{\labelsep}{0.5em} } 
}

\newcommand{\squishend}{
  \end{list} 
}

\author{Simos Gerasimou}
\email{simos.gerasimou@york.ac.uk}
\affiliation{%
	\institution{University of York}
	\city{York}
	\state{UK}
}

\author{Hasan Ferit Eniser}
\email{hfeniser@mpi-sws.org}
\affiliation{%
	\institution{MPI-SWS}
 	\authornote{Work done while at Bogazici University.}
	\city{Kaiserslautern}
	\state{Germany}
}
 \authornotemark[1]

\author{Alper Sen}
\email{alper.sen@boun.edu.tr}
\affiliation{%
	\institution{Bogazici University}
	\city{Istanbul}
	\state{Turkey}
}

\author{Alper Cakan}
\email{alper.cakan@boun.edu.tr}
\affiliation{%
	\institution{Bogazici University}
	\city{Istanbul}
	\state{Turkey}
}

\definecolor{greenCode}{RGB}{153, 213, 202}
\definecolor{blueCode}{RGB}{131, 187, 229}
\definecolor{yellowCode}{RGB}{252,196,56}

\newcommand{\approach}{DeepImportance}

\algnewcommand{\IIf}[1]{\State\algorithmicif\ #1\ \algorithmicthen}
\algnewcommand{\EElse}[1]{\State\algorithmicelse\ #1\ }
\algnewcommand{\EndIIf}{\unskip}

\setcopyright{acmcopyright}
\copyrightyear{2020}
\acmYear{2020}
\acmDOI{10.1145/1122445.1122456}

\acmConference[ICSE '20]{ICSE '20: The 42th International Conference on Software Engineering}{May 23--29, 2020}{Seoul, South Korea}
\acmBooktitle{ICSE '20: Proceedings of the 42th International Conference on Software Engineering, May 23--29, 2020, Seoul, South Korea}
\acmPrice{00.00}
\acmISBN{978-1-4503-9999-9/18/06}

\acmSubmissionID{123-A56-BU3}

\begin{document}

%\title{Guiding Deep Learning System Testing\\ using Relevance Propagation}
\title{
%	How Important is your Neuron?\\
%	Important-Guided Deep \\	Learning System Testing}
Importance-Driven Deep Learning System Testing}

\begin{abstract}
Deep Learning (DL) systems %have been characterised as 
are key enablers for engineering intelligent applications due to their ability to solve complex tasks such as image recognition and machine translation. 
Nevertheless, using DL systems in safety- and security-critical applications requires to provide testing evidence for their dependable operation. 
Recent research in this direction focuses on adapting testing criteria from traditional software engineering as a means of increasing confidence for their correct behaviour. 
However, they are inadequate in capturing the intrinsic properties exhibited by these systems.
We bridge this gap by introducing \approach, a systematic testing methodology accompanied by an \textit{Importance-Driven} (IDC) test adequacy criterion for DL systems.
%proposing a novel test adequacy criterion for DL systems called Importance-Driven Coverage (IDC). 
Applying IDC enables to establish a \textit{layer-wise functional understanding} of
the importance of DL system components 
%by analysing its behaviour under a given data 
and use this information to 
assess the \textit{semantic diversity} of a test set.
%guide the generation of \textit{semantically-diverse} test sets.
%using the calculated importance scores to guide test selection and retraining. 
Our empirical evaluation on several DL systems, across multiple DL datasets and with state-of-the-art adversarial generation techniques demonstrates the usefulness and effectiveness of \approach\ and its ability to support the engineering of more robust DL systems.

%autonomous applications including driverless vehicles and socially assistive robots. 
%Nevertheless, their use in safety and security critical applications requires these systems to be thoroughly tested. 

%that require image recognition and speech translation. Nevertheless, their use in safety and security critical applications, 

%developing autonomous systems in including driverless vehicles and socially assistive robots. 

\end{abstract}

%ALPER: These codes need to be generated automatically from the following site.

% at https://dl.acm.org/ccs/ccs.cfm




\maketitle

% !TEX root = ../main.tex

\section{Introduction}
\label{sec:introduction}
\noindent

Driven by the increasing availability of publicly-accessible data and massive parallel processing power, Deep Learning (DL) systems have achieved unprecedented progress, commensurate with the cognitive abilities of humans~\cite{LecunBH2015,Goodfellow-et-al-2016}. 
In fact, DL systems can solve challenging real-world tasks such as image   classification~\cite{CiresanMS2012}, natural language processing~\cite{SutskeverVQ2014} and speech recognition~\cite{HintonDYD2012}. Consequently, DL systems are becoming key enablers in many applications, including medical diagnostics~\cite{litjens2017survey}, air traffic control~\cite{julian2016policy},  malicious code detection~\cite{CuiXCCWC2018} and autonomous vehicles~\cite{bojarski2016end}.

Despite the manifold potential applications, using DL systems in safety- and security-critical applications requires the provision of assurance evidence for their trustworthy and robust behaviour~\cite{BurtonGH2017}.
%Defects and 
Vulnerabilities and defects in these systems, either originating from systematic errors, insufficient generalisation or inadequate training, can endanger human lives, lead to environmental damage or cause significant financial loss~\cite{varshney2016engineering}. 
Preliminary reports from safety advisory boards (e.g., the US transportation board~\cite{NTSB}) regarding recent unfortunate events involving autonomous vehicles~\cite{Google,Uber} underline not only the challenges associated with using DL systems but also %and emphasise 
the urgent need for improved assurance evaluation practices.
%their reliable assessment.
%providing quality assurance guarantees. 

%\textbf{Motivation + Examples}
From a safety assurance perspective, testing has been among the primary instruments for evaluating quality properties of software systems providing a trade off between completeness and efficiency~\cite{jorgensen2013software}. 
%For instance, 
Domain-specific standards such as ISO26262~\cite{ISO26262} and DO-178C~\cite{DO178} prescribe testing principles (e.g., adequacy criteria, testing properties) which should be employed for the verification of applications within the automotive and avionics domains, respectively. 
Evidence collected as a result of testing is typically used to 
%The correct execution of testing activities provides the evidence required for 
demonstrate compliance with 
%establishing the 
expected quality assurance levels, thus manifesting 
%that hazardous behaviours have been mitigated and 
the ability of those systems to operate with an acceptable risk of failure within their lifetime. 

However, testing DL systems by simply adopting principles recommended by these standards is not straightforward~\cite{SHK18,borg2018safely}.
%Testing DL systems by directly adopting principles from these standards presents several challenges~\cite{SHK18}.
The lack of a system specification regulating the inference mechanism to be learnt combined with the data-driven programming paradigm makes impossible to explicitly encode the expected DL system behaviour into its control flow structures~\cite{salay2018analysis}.
The extremely large configuration spaces of modern DL models deteriorates the issue as it is impossible to determine and calibrate the influence of each configurable parameter in completing a task; e.g., LeNet~\cite{lecun1998mnist} and VGG-16~\cite{simonyan2014very} have more than 60K and 100M configurable parameters, respectively. 
%Even worse, determining the influence of each configurable parameter in completing a task is very challenging, especially due to the extremely large configuration spaces of modern DL models; e.g., LeNet~\cite{lecun1998mnist}, VGG-16~\cite{simonyan2014very} have more than 60K and 100m parameters, respectively. 
Thus, traditional software testing techniques and %structural 
coverage criteria~\cite{ammann2016introduction} 
%cannot support DL system verification.
are inapplicable for DL system verification. 

%manual ad hoc testing of DL systems

%\textbf{State-of-the-art}\textbf{Limitations}
Driven by the need for providing high-quality assurance in DL systems
%for systematic and effective testing for DL systems 
and inspired by traditional software engineering testing paradigms~\cite{ammann2016introduction}, recent research proposes novel testing techniques and coverage criteria~\cite{OG18,Kim2019aa, pei2017deepxplore,MJX18,SWRHKK18} (see Section~\ref{sec:relatedWork} for an overview of related work).
%from traditional software engineering to DL systems~\cite{pei2017deepxplore,MJX18,SWRHKK18,OG18,Kim2019aa}.
The core principle underlying those techniques is that for effective DL system testing, the test set should be characterised by high diversity, thus enabling to exercise different behaviours of the system~\cite{MJX18}. 
For example, DeepXplore~\cite{pei2017deepxplore} estimates the diverse DL system behaviour by calculating \textit{neuron coverage} as the ratio of neurons whose activation values are above a predefined threshold. 
Similarly, the DeepGauge multi-granularity testing criteria~\cite{MJX18} generalise the neuron coverage concept and 
calculate the ability of the test set to cover (i.e., trigger) major and corner-case neuron regions, given by partitioning the ranges of neuron activation values. 
Despite their usefulness, these criteria are simply an aggregation of neurons (or neuron regions) whose activation values conform to certain conditions. 
%In fact, the inability to provide insi	ghts regarding the improvement contributed by individual test inputs leads to a weak causal relationship which  
By focusing only on these constrained neuron properties and ignoring the overall DL system behaviour, the causal relationship between the test set and decision-making is uninformative~\cite{Kim2019aa}. % and weak. 
Also, the instantiation of recently proposed techniques depends on user-defined conditions (number of regions~\cite{MJX18} or upper bounds~\cite{Kim2019aa}) which might not represent the actual behaviour of the DL system adequately. 
%typically the instantiation of the required to instantiate these techniques are typically 
%\textcolor{blue}{On the other side, the existing criteria counting unexpected neuron activation traces either in a layer-wise manner \cite{OG18,Kim2019aa} or in the whole network \cite{Kim2019aa} can not provide an insight on adequacy of a given test set without a user defined condition. In addition to that, none of the existing works can capture the functional behaviour of a DL system.}
Finally, these criteria provide limited information about the testing improvement contributed by individual test inputs as expected for an effective testing adequacy criterion~\cite{goodenough1975toward,ammann2016introduction}.
%Consequently, the causal relationship inferred by these criteria between the test input and DL system behaviour is weak. 
%Thus, they provide a weak causality analysis between the  test inputs and correct classifications/misclassifications

In this paper, we bridge the gap in existing research by introducing 
\textit{\approach}, a systematic testing methodology accompanied by an \textit{Importance-Driven} test adequacy criterion for DL systems based on relevance propagation. 
%At a high-level, 
%In particular, 
By analysing the activity of a DL system and its internal neuron behaviours, \approach\ develops a \textit{layer-wise functional understanding} that signifies the contribution of internal neurons to the output through the layers. 
This contribution enables to determine the causal relationship between the neurons and the DL system behaviour as more influential neurons have a stronger causal relationship and can \textit{explain} which high-level features influence more the decision-making. 
%are involved in decision of the DL system
%The most influential neurons explains
%\textcolor{blue}{In other words, the most relevant neurons i.e., those with the highest relevance scores, explain which high-level features are involved in decision of the DL system.}
\approach\ establishes this relationship by computing a decomposition of the decision made by the DL system and iteratively redistributing the relevance in a layer-wise manner proportional to how prominent each neuron and its connections are~\cite{bach2015pixel}. 
As we demonstrate in Section~\ref{ssec:neuronAnalysis}, this importance score is quite different from the neuron activation values used by similar DL testing techniques (e.g.,~\cite{MJX18,pei2017deepxplore}).
Using those important neurons, \approach\ carries out \textit{neuron-wise quantisation} to partition the space of each neuron's activity into an automatically-determined finite set of clusters that captures its behaviour to a sufficient level of granularity. 
%\approach\ calculates the adequacy of an input set using the \textit{Importance-Driven} adequacy criterion given by the ratio of combinations of important neurons clusters covered by the set.
The \textit{Importance-Driven} adequacy criterion instrumented by \approach\ measures the adequacy of an input set as the ratio of combinations of important neurons clusters covered by the set.
%\textcolor{blue}{
%\approach\ calculates the adequacy of a set of test inputs as the percentage of the possible combinations of quantised neuron clusters covered by the set.}
%The layer-based \approach-driven result of a set of test inputs is calculated as the percentage of the possible combinations of quantised neuron clusters  covered by the set. 

Our empirical evaluation using publicly available datasets (MNIST \cite{lecun1998mnist}, CIFAR-10~\cite{cifar_model}, Udacity self-driving challenge~\cite{udacity}) and DL systems whose models size ranges from small-medium (LeNet~\cite{lecun1998mnist}) to large (e.g., Dave-2~\cite{bojarski2016end}) %for autonomous vehicles) 
demonstrates the ability of \approach\ to develop a functional understanding of the DL system and evaluate the testing adequacy of a test set. 
Furthermore, the \textit{Importance-Driven} adequacy criterion is effective  in quantifying the ability of a DL system to identify defects as indicated by the coverage difference between the original test set and adversarial examples generated using state-of-the-art adversarial generation techniques~\cite{goodfellow2015explaining,papernot2016limitations,carlini2017towards,kurakin2016adversarial}.
%which is in line with existing research in DL system testing~\cite{pei2017deepxplore,MJX18}.
%effective in quantifying the defect

%\textbf{Contributions}
Overall, the main contributions of our paper are:
\squishlist
	\item The \approach\ approach for finding important neurons of a DL system that are core contributors in decision-making; 
	%and thus guiding allocation of testing resources; %(i.e., they should be tested more extensively);
	\item The \textit{Importance-Driven Coverage} criterion which can establish the adequacy of an input set to trigger different combinations of important neurons' behaviours, thus enabling software engineers to assess the semantic adequacy of a test set; 
	\item Am extensive \approach\ evaluation on three public datasets (MNIST, CIFAR-10, Udacity) and three DL systems (LeNet, CIFAR-10, Dave-2) showing its feasibility and effectiveness;
	\item A prototype open-source \approach\ tool and a repository of case studies, both of which are freely available from our project webpage at \url{https://deepimportance.github.io}.
\squishend

To the best of our knowledge, \approach\ is the first systematic and automated testing methodology that employs the semantics of neuron influence to the DL system as a means of developing a laywer-wise functional understanding of its internal behaviour and assessing the semantic adequacy of a test set. 
%evaluating the adequacy of a test set and guiding test synthesis. 

% \textbf{Structure}
The remainder of the paper is structured as follows.
Section~\ref{sec:background} presents briefly DL systems and coverage criteria in traditional software testing. 
Section~\ref{sec:approach} introduces \approach\ and Section~\ref{sec:implementation} presents its open-source implementation. Section~\ref{sec:evaluation} describes the experimental setup %, esearch questions 
and evaluation carried out.
Sections~\ref{sec:relatedWork} and~\ref{sec:conclusion} discuss related work and conclude the paper, respectively. 

% !TEX root = ../main.tex

\begin{figure}[t]
	\centering
	\includegraphics[width=0.8\linewidth]{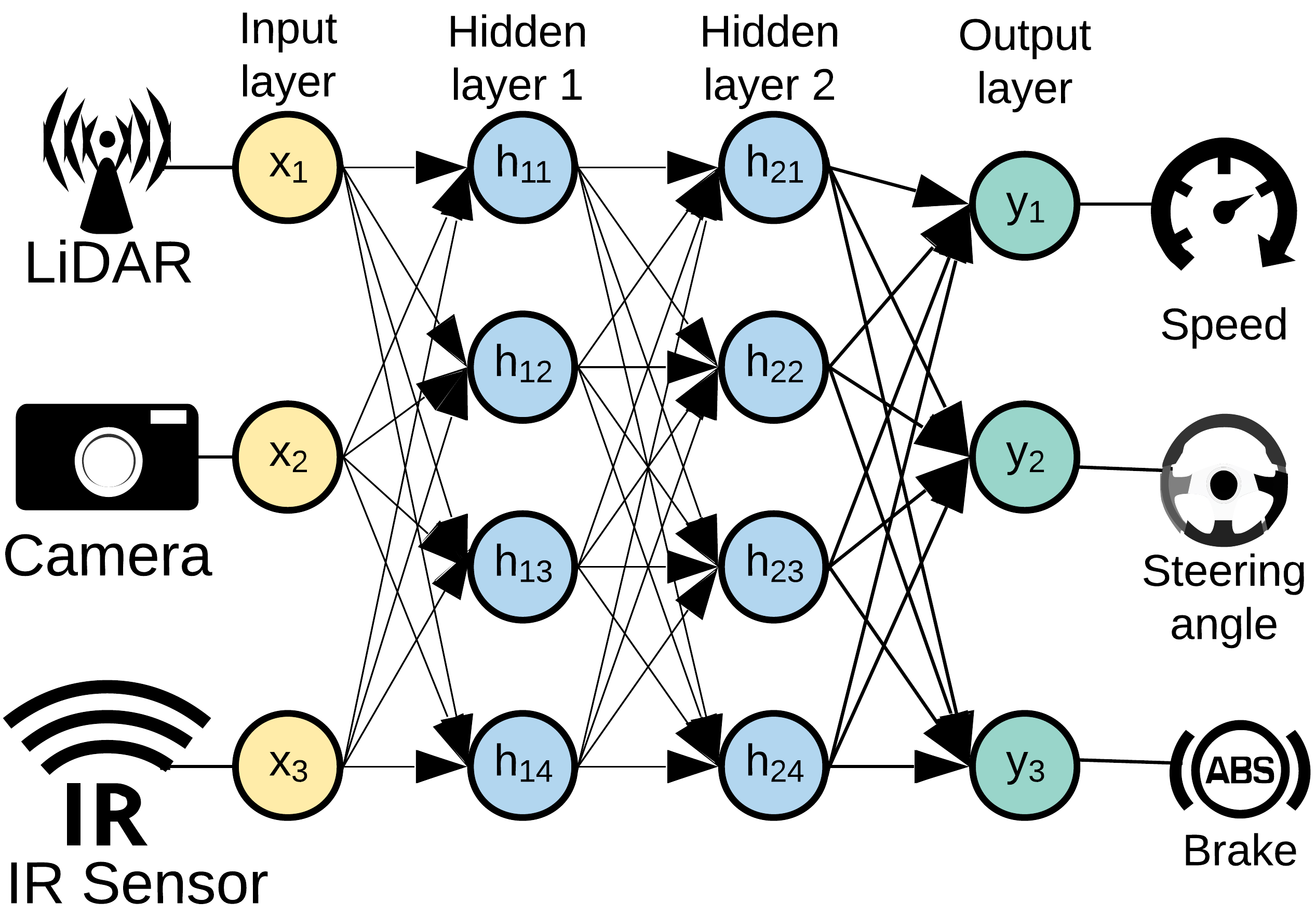}
	
%	\vspace*{-3mm}	
	\caption{A four layer fully-connected DL system that receives inputs from vehicle sensors (camera, LiDAR, infrared) and outputs a decision for speed, 
		steering angle and brake.}
	\label{fig:dnn}
	
%	\vspace*{-4mm}
\end{figure}

\begin{figure*}[t]
	\centering
	\includegraphics[width=\linewidth]{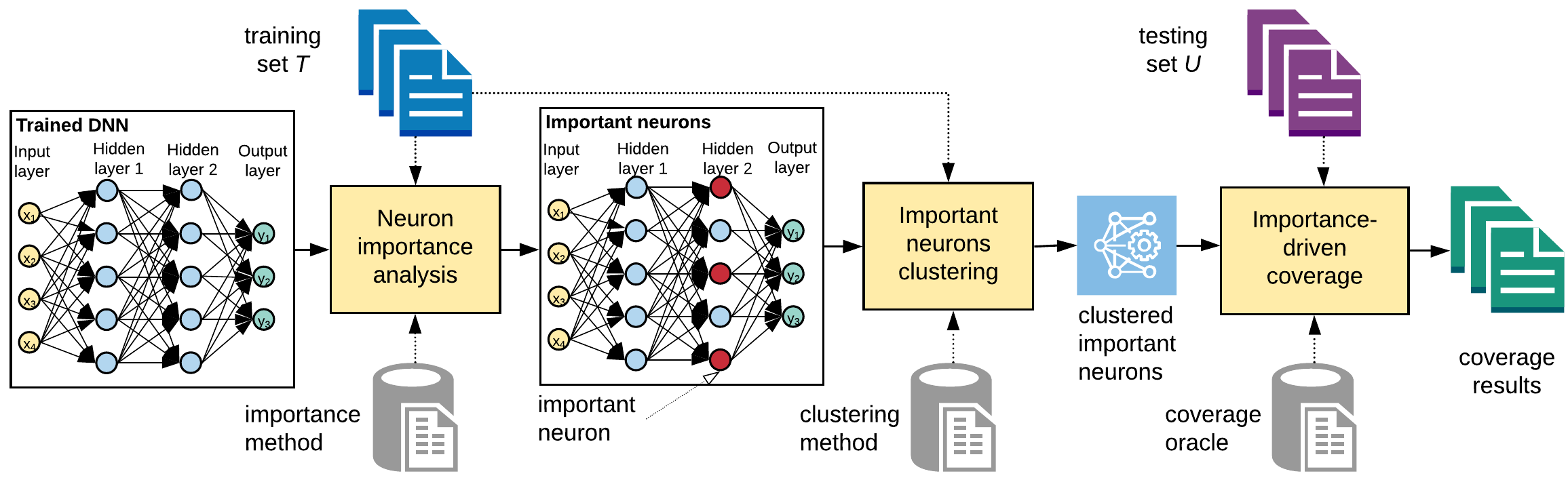}
	
%	\vspace*{-2mm}
	\caption{\approach\ workflow for determining the importance-based testing adequacy of a DL system.}
	
	\label{fig:overview}
%		\vspace*{-4mm}
\end{figure*}

\section{Background}
\label{sec:background}
\noindent

\vspace*{-2mm}
\subsection{Deep Learning Systems \label{ssec:dnns}}

Fig. \ref{fig:dnn} shows a typical feed-forward DL system consisting of several interconnected neurons arranged into consecutive layers: the input layer, the output layer and at least one hidden layer~\cite{Goodfellow-et-al-2016}. 
Each layer within a DL system comprises a sequence of neurons. 
A neuron represents  a computing unit that applies a nonlinear activation function to its inputs and transmits the result to neurons in the following layer~\cite{LecunBH2015}. 
Commonly used activation functions include sigmoid, hyperbolic tangent and ReLU (Rectified Linear Unit). 
All neurons, except from those in the input layer, are connected to neurons in the following layer with weights whose values express how strong are the connections among neuron pairs.
A DL system's architecture comprises the number of layers, neurons per layer, neuron activation functions and a cost function.
Given such an architecture, the DL system carries out an iterative training process through which it 
consumes %a large amount of 
labelled input data (e.g, raw image pixels) in its input layer, 
executes a set of nonlinear transformations in its hidden layers to extract semantic concepts (i.e., features) from the input data, 
and, finally, generates a decision that matches the effect of these computations in its output layer.
The training process aims at finding
%The aim of this training process is to find 
weight values that minimize the cost function, thus enabling the DL system to achieve high generalisability.

\subsection{Coverage Criteria in Software Testing \label{ssec:coverage}}
Since testing a software system exhaustively is, in principle, impossible due to its extremely large number of possible inputs, coverage criteria are typically employed to quantify how well a test suite exercises the system~\cite{ammann2016introduction,pezze2008software}.
There are several types of coverage criteria, with the most widely-adopted in industry being: statement coverage, condition coverage, path coverage and branch coverage~\cite{jorgensen2013software}. 	

Testing techniques and coverage criteria are building blocks of safety standards employed in various safety-critical domains such as automotive and avionics (e.g.,  ISO26262~\cite{ISO26262}, DO-178C~\cite{DO178}). 
Depending on the integrity level associated with a system component, different coverage criteria are mandated. 
For instance, ISO26262 requires to demonstrate compliance with statement coverage and MC/DC (Modified Condition/Decision coverage) for components whose integrity levels are the lowest (A) and highest (D), respectively. 
The higher the risk from a component's misbehaviour, the higher its integrity level, and thus, more significant assurance (testing) effort is required to avoid unreasonable residual risk.

% !TEX root = ../main.tex

\section{Approach}
\label{sec:approach}

\approach, whose high-level workflow is shown in Fig.~\ref{fig:overview}, enables the systematic testing and evaluation of DL systems. 
%In this section, we introduce the \approach\ approach that enables the systematic testing and evaluation of DL systems. 
%The high-level \approach\ workflow is shown in Figure~\ref{fig:overview}. 
Using a pre-trained DL system, \approach\ analyses the training set $\mathcal{T}$ to establish a fundamental understanding of the overall contribution made by internal neurons of the DL system. 
This enables to identify the most important neurons that are core contributors to the decision-making process (Section~\ref{ssec:neuronAnalysis}). 
%network towards achieving the current performance level of the DL system 
Then, \approach\ carries out a quantisation step 
which produces an \textit{automatically-determined} finite set of clusters of neuron activation values that characterises, to a sufficient level, how the behaviour of the most important neurons changes with respect to inputs from the training set (Section~\ref{ssec:neuronQuantisation}). 
Finally, \approach\ uses the produced clusters of the most important neurons to assess the coverage adequacy of the test set%$U$
~(Section~\ref{ssec:relevanceCoverage}).
Informally, the \textit{Importance-Driven} test adequacy criterion 
of \approach\ 
is satisfied when all combinations of important neurons clusters are exercised.

We use the following notations to present \approach.  
Let $D$ be a DL system with $L$ layers. Each layer $L_i,  1 \leq i \leq L$, comprises $|L_i|$ neurons and the total number of neurons in $D$ is $S = \sum_{i=1}^{L}|L_i|$. Let also $n_{i,j}$ be the $j$-th neuron in the $i$-th layer. When the context is clear, we use $n \in D$ to denote any neuron that is a member of $D$ irrespective of its layer. 
%ALPER: Do we need D_H?
%Similarly, we use $D_H$ to denote the neurons which belong to the hidden layers of $D$, i.e., $D_H= \{n_{ij} | 1 \leq i < L, 1 \leq j \leq |L_i| \}$.
%ALPER: what is small x? a set of ...
Let $X$ denote the input domain of $D$ and $x \in X$ be a concrete input. 
%and $u \in x$ for an element of $x$.
%Let $X = \{x_1, x_2, ...\}$ be a set of inputs. 
%We use $T$ to denote the set of test inputs from the input domain of $D$, $t \in T$ to denote a concrete input, and $u \in t$ for an element of $t$. 
Finally, we use the function $\phi(x, n) \in \mathbb{R}$ to signify the output of the activation function of neuron $n \in D$.

\subsection{Neuron Importance Analysis \label{ssec:neuronAnalysis}}

%The identification of neurons within a DL system that are key contributors to the system's performance 

The purpose of \textit{importance analysis} is to identify neurons within a DL system that are key contributors to decision-making. %execution. 
Given an input, information within a DL system is propagated according to the strength of connections (weights) between neurons in successive layers.
As such, the activity of some neurons influences more the capabilities of the system to make correct decisions~\cite{LecunBH2015}.

Although \textit{representation learning} is a key characteristic of DL systems that eliminates the tedious and potentially erroneous process of manual feature extraction, it also means that neurons develop, through backpropagation~\cite{rumelhart1986learning},  the ability to learn optimal feature transformations for the given setting on their own~\cite{bengio2013representation}.
%identify specific features for a given class of inputs
More specifically, raw input data passing through the complex architecture of a modern DL system, with many layers, many neurons per layer and non-linear  transformations (e.g., ReLU activation functions~\cite{maas2013rectifier}, max pooling, convolutions), yield abstract and discriminative features that enable the system to make effective decisions in the final layer using a log-linear model (typically softmax)~\cite{LecunBH2015}. 
For instance, neurons within the initial hidden layers learn abstract shapes (e.g., edges, circles) while neurons in deeper layers extract more semantically meaningful features (e.g., faces, objects).
Using as an analogy a software system whose architecture adopts 
conventional software engineering principles, neurons can be considered as functions that execute a distinct functionality. Irrespective of the position of a function into the control flow graph, it receives (transformed) information from functions preceding in the graph and itself applies function-specific transformations before propagating the updated information to
subsequent functions in the control flow graph.
% other subsequent functions.  

We capitalise on this unique characteristic of neurons within a trained DL system to establish the importance of each neuron. 
To achieve this, we compute a decomposition of the decision $f(x)$ made by the system for input $x \in X$ and use \textit{layer-wise relevance propagation}~\cite{bach2015pixel} to traverse the network graph and redistribute the decision value in a layer-wise manner proportional to the contribution made by each neuron within the layer. 
%The overall relevance of each neuron in the lower layer is determined by summing up the relevance coming from all upper-layer neurons.
For a fully-connected layer $i$, the relevance $R_{ij}$ of the $j$-th neuron entails redistributing relevance from neurons in layer $i+1$ which is given by~\cite{bach2015pixel}:
%to neurons in layer $l$ 
\begin{equation}\label{eq:redistributionRule}
R_{ij} = \sum_k \frac{\phi(x,n_{ij}) w _{ijk}}{\sum_i \phi(x,n_{ij}) w_{ijk}+\epsilon} R_{i+1,k}
\end{equation}

\noindent
where $R_{i+1,k}$ is the relevance score of the $k$-the neuron in layer $i+1$, $w_{ijk}$ is the weight connecting neuron $j$ to neuron $k$ and $\epsilon$ is a small stabilization term (to avoid division by zero). 

\begin{algorithm}[t]
	\caption{Neuron importance analysis}\label{alg:analysis}
	\begin{small}
		\renewcommand{\baselinestretch}{1}
		\begin{algorithmic}[1]
			\Function{ImportantNeuronsAnalysis}{$\mathcal{D}, \mathcal{X}, m$}
			\State $R \gets \emptyset$\label{a1:l1}
			\hfill\Comment relevant neurons vector
			\ForAll{$x\in\mathcal{X}$}\label{a1:allX}
			\State $R^x \gets \emptyset$\label{a1:l1}\Comment $x \in \mathcal{X}$ importance vector				
			\State $R_{L} = \textsc{ComputeValue}(\mathcal{D}, x)$ \hfill\Comment decision value
			\ForAll{$i \in \{L_{L-1}, ..., L_1\}$} \Comment relevance propagation
			\State $R_{L}\!\!=\!\! \textsc{Relevance}(L_i, x, R_L)$ \hfill\Comment $L_i$ neurons relevance
			\State $R^x = R^x \frown R_{L}$ \Comment append to vector $R^x$
			\EndFor
			\State $R = R \cup R^x$ \Comment collect relevance vectors
			\EndFor
			\State $AN =\textsc{Analyse}(D,R)$ \Comment analyse relevance vectors
			%			\State $RN \!=\!\! \{n | n \in \textsc{Analyse}(D,R)\}$ \Comment select top $k$ neurons
			\State \Return $\textsc{Top}(AN, m)$ \Comment select top $m$ neurons
			\EndFunction
			
			%		\vspace*{-2mm}
		\end{algorithmic}
	\end{small}
	
\end{algorithm}

Intuitively, the relevance attributed to neurons in layer $i$ from neurons in layer $i+1$ is proportional to (i) the neuron activation $\phi(x,n_{ij})$, i.e., neurons with higher activation values receive a larger relevance contribution; and (ii) the strength of the connection $w_{ijk}$, i.e., more relevance flows through more important connections. 
While (\ref{eq:redistributionRule}) applies to fully-connected layers, we refer interested readers to~\cite{montavon2017explaining} for definitions of redistribution rules for other layer types including pooling, activation and normalisation layers. 

The redistribution process is underpinned by a \textit{relevance conservation} property specifying that at every step of the process (i.e., at every layer $L_i$) the total amount of relevance (i.e., the prediction) is conserved. No relevance is artificially added or removed.\footnote{When neurons with bias contribute to the output, the relevance attribution to the bias is redistributed onto each input of the decomposed layer using the method in~\cite{montavon2017explaining}.}.
Therefore, $\sum_j^{|L_1|} R_{1j} = \cdots = \sum_k^{|L_4|} R_{4k} = \sum_{l}^{|L_5|} R_{5l} = \cdots = f(x)$.
\noindent

Algorithm~\ref{alg:analysis} shows the high-level process for computing the importance scores for neurons of $D$ and selecting the $m$ most important. For any given input $x \in X$, we perform a standard forward pass to compute the decision value, i.e., the magnitude of evidence for a given class before applying softmax (line~5). 
Next,  we perform a backward pass (lines 6--8) considering each layer successively during which the relevance is allocated to neurons of the current layer before being backpropagated from one layer to another until it reaches the input layer.
The decomposition is achieved using the layer-specific rules in~\cite{bach2015pixel}. The \textsc{Analyse} function (line 10) analyses the relevance scores of all neurons %\sg{in the target layer} 
for all inputs and prioritises them based on a priority criterion (e.g., cumulative relevance, normalised relevance). In our evaluation (Section~\ref{sec:evaluation}), we use cumulative relevance.
Finally, the top $m$ neurons are returned (line 11). 
%\sg{Should we say a few words here about the selection of $m$?}

The use of relevance for identifying the most important neurons is a key ingredient of our approach. 
%Differently though from recent research on \textit{explainability} of DL systems, which targets the identification of input parts responsible for the prediction, \approach\ targets the 
%Nevertheless, other conserving backward propagation techniques (e.g., DeepLift) could be used. 
Building on recent research on \textit{explainability} of DL systems,  which targets the identification of input parts responsible for a prediction, \approach\ targets the identification of the most influential neurons; these are \textit{high-risk neurons} that should be tested thoroughly. 
Albeit being outside the scope of this work, we also highlight that other explainability- driven techniques could be used for the identification of the most important neurons (e.g., DeepLift~\cite{shrikumar2017learning}, L2X~\cite{chen2018learning}).
%those high risk neurons 
%\cite{Sundararajan2017,chen2018learning}

\begin{figure}
	\centering
	\includegraphics[height=3cm]{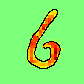}
	\includegraphics[width=5cm]{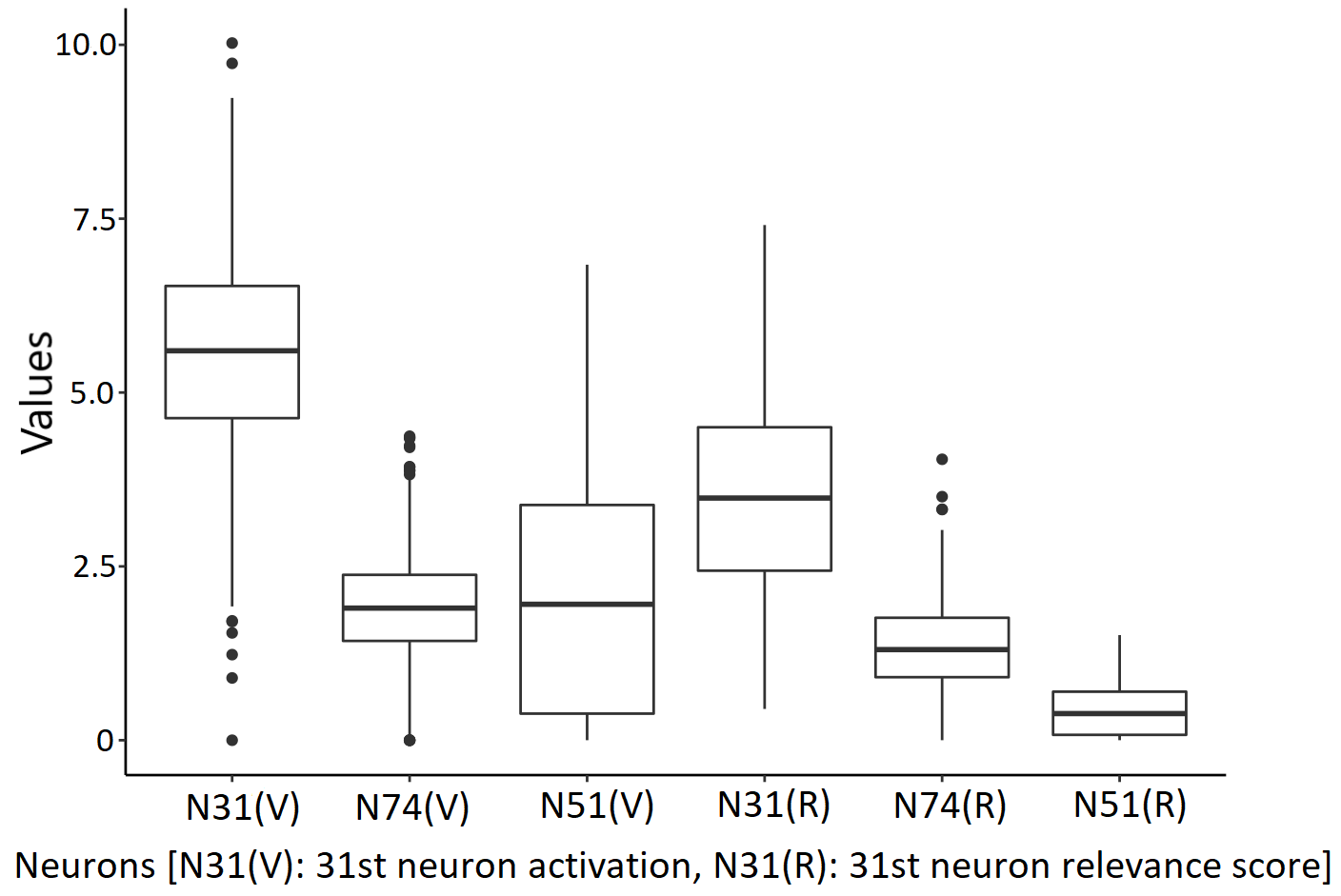}
	
	\vspace*{-2mm}	
	\caption{Input from MNIST dataset with the most important pixels contributing to the correct decision highlighted (left), and difference between neuron activation values and relevance scores (right) for the same set of neurons.}
	
	\vspace*{-2mm}	
\label{fig:lrpM}
\end{figure}

State-of-the-art testing adequacy criteria for DL systems including neuron coverage~\cite{pei2017deepxplore} and k-multisection neuron coverage~\cite{MJX18} quantify testing coverage solely based on neuron values, 
irrespective of the added value of a neuron  to the final decision. 
In other words, a neuron might contribute to increasing the confidence for classes other than the correct one, and this is not distinguished.
\approach\ captures the actual contribution made by each neuron to the decision which in shallow and deeper layers corresponds to raw pixels and concrete features from the input domain, respectively. 
For instance, Fig.~\ref{fig:lrpM}~(left) shows the most important pixels and Fig.~\ref{fig:lrpM}~(right) shows the difference between the activation values and the relevance scores for the same set of most important neurons within the penultimate layer of a LeNet network~\cite{lecun1998mnist}.
\approach\ exploits this understanding to assess the adequacy of a test set to examine the most critical neurons, i.e., those with the strongest influence on the behaviour of the DL system. 
%(cf. robustness results in Section~\ref{ssec:results}). 
%guide testing towards 

Using relevance is also significantly different compared to sensitivity analysis~\cite{montavon2017explaining}. 
While sensitivity analysis cares about \textit{what makes more/less a labelled input (e.g., a dog) to be classified as its target label}, relevance analysis investigates \textit{what actually makes the input to be classified as that label}. 
The sensitivity scores do not really explain why an input has been predicted in a certain way, but rather to which direction in the input space the output is most sensitive. 
In contrast, relevance scores indicate which neurons/inputs are pivotal for the classification.
Thus, they are a significantly more informative and practicable measure for assessing and explaining the composition about the decision made by the DL systems~\cite{bach2015pixel}.

\subsection{Important Neurons Clustering\label{ssec:neuronQuantisation}}
Having established the important neurons 
that are core contributors to the behaviour of the DL system, we are now ready to determine regions within their value domain which are central to the DL system execution. 
Since each neuron is responsible for perceiving specific features within the input domain~\cite{LecunBH2015}, we argue that for inputs with similar features the activation values of those important neurons are concentrated into specific regions within their value domain. 
Informally, those regions form a pattern that captures the activity of the most influential neurons of the DL system.

The purpose of clustering is threefold. 
First, compared to~\cite{MJX18} which partitions the value range of neuron activation values into $k$ buckets of equal width solely based on a randomly selected number of buckets (i.e., $k$-multisection neuron coverage~\cite{MJX18}), the clusters generated by our approach correspond to \textit{semantically different features} of each neuron.
%, thus informing the DL system testing. 
%\sg{Do we show this?}
Second, since the range of neuron activation values $\phi(x,n)$ could in principle be the entire set of real numbers ($\mathbb{R}_+$ for ReLU activation functions), the cyclomatic complexity for analysing the DL system is very large. 
Similar to techniques employed in~\cite{Kim2019aa,MJX18}, clustering (bucketing in~\cite{Kim2019aa,MJX18}) enables to reduce dimensionality and computational cost, thus making tractable to test the DL system (cf. Section~\ref{sec:evaluation}). 
Finally, the identification of clusters for those important neurons could inform the allocation of testing resources to ensure that the regions of those neurons are tested sufficiently, thus increasing our confidence for the robust DL system behaviour. 
%and, as described in Section~\ref{ssec:relevanceCoverage}, dedicating more resources to test sufficiently those regions is important to increase our confidence for the correct DL system behaviour. 

\approach\ employs \textit{iterative unsupervised learning} to cluster the vector of activation values from the training set for each important neuron and determine sets of values that can be grouped together. 
%by employing \textit{unsupervised learning} to cluster clustering the vector of activation values for each important neuron given the training set $\mathcal{T}$.  
The \approach\ instantiation we present in this research work (Section~\ref{sec:evaluation}) employs \textit{k-means}~\cite{kmeans}, an iterative clustering method that produces $k$ clusters which minimize the within-class sum of squares. 
To this end, we segment the activation values of each important neuron into groups (clusters) so that activation values within the same group are more similar to other activation values in the same group and dissimilar to those in other groups. 
%%pairwise dissimilarities between activation values in the same group are smaller than those in different groups. 

Determining the optimal number of clusters without analysing the data is not a trivial problem~\cite{kaufman2009finding}.
We reinforce cluster extraction with the \textit{Silhouette index}~\cite{rousseeuw1987silhouettes}, thus supporting the \textit{automatic identification} of a neuron-specific optimal strategy for clustering the activation values of each important neuron in $D_m$. 
Silhouette is an internal clustering validation index that computes the goodness of a clustering structure without external information~\cite{liu2010understanding}.
As such, depending on each neuron's activation values, the optimal number of clusters is determined automatically and can be different between the important neurons.
Also, this strategy addresses the weakness of k-means that requires to define the desired number of clusters a priori. 
More formally, given the $n$-th important neuron, $n \in D_m$, and 
the function $C(t)$ indicating for each $t \in T$ the cluster assigned to $\phi(t, n)$ within the $n$-th neuron's clusters,
%a training set $T$ such that $\phi(t, n) \in C(t)$ for each $t \in T$,
the Silhouette score for $c \in \mathbb{N_+}$ clusters is defined as follows

\begin{equation}\label{eq:silhouette}
S_n^c = \frac{1}{|T|}\sum_{t=1}^{|T|} \frac{B(t)-A(t)}{max(B(t), A(t))}
\end{equation}
where

\begin{equation}\label{eq:silhouetteA}
%A(t) = \frac{1}{1-|C_c|} \sum_{\phi(u,n) \in \{C_c \setminus \phi(t,n) \}} d(\phi(u,n), \phi(t,n))
A(t) = \frac{1}{1-|C(t)|} \sum_{\phi(u,n) \in \{C(t) \setminus \phi(t,n) \}} d(\phi(u,n), \phi(t,n))
\end{equation}

\noindent is the \textit{intra-cluster cohesion}  given by the average $L_1$ distance of activation value $\phi(t,n)$ to all other values in the same cluster, and  
\begin{equation}\label{eq:silhouetteB}
%B(t) = \min_{C_k \neq C_c}\frac{1}{|C_k|}\sum_{\phi(u,n) \in C_k} d(\phi(u,n), \phi(t,n))
B(t) = \min_{C' \neq C(t)}\frac{1}{|C'|}\sum_{\phi(u,n) \in C'} d(\phi(u,n), \phi(t,n))
\end{equation}

\noindent is the \textit{inter-cluster separation} given by the average $L_1$ distance between $\phi(t,n)$ and activation values in its nearest neighbour cluster.

Maximising the Silhouette score gives the optimal clustering strategy and correspondingly the optimal number of clusters for the $n$-th important neuron.
Therefore, 
the higher the score the better the overall quality of the clustering result in terms of cluster cohesion and cluster separation.

\algdef{SE}[DOWHILE]{Do}{doWhile}{\algorithmicdo}[1]{\algorithmicwhile\ #1}%

\begin{algorithm}[t]
	\caption{Important Neurons Cluster Extraction}\label{alg:quantisation}
	\begin{small}
		\renewcommand{\baselinestretch}{1}
		\begin{algorithmic}[1]
			\Function{ClusterImportantNeurons}{$\mathcal{D}_{m}, \mathcal{T}, \mathcal{C}$}
			\State $\Psi \gets \emptyset$\label{a2:l1}
			\hfill\Comment vector for clustered important neurons
			\ForAll{$n\in\mathcal{D}_{m}$}\label{a2:allN}
			\State $\Phi_n = (\phi(t,n)), t \in \mathcal{T}$ \Comment $n$-th neuron activation values\label{al2:PhiN}
			%	  			\Do
			%				\State $\tesc{Get}Assign
			%				\doWhile{$\;\lnot\textsc{Terminate}(Q_n)$}				
			%				\While {$\;\lnot\textsc{Terminate}(Q_n)$}
			%				\EndWhile 
			%			\State $L_n = (\textsc{Labels}(\Phi_n, c)), c\in C$
			%			\State $c^\text{min}_n = \text{arg}\; \text{min}_{c \in C}\;\textsc{Score}(\Phi_n, \textsc{GetLabels}(L_n, c))$
			\State $c^\text{max}_n = \text{arg}\; \text{max}_{c \in C}\;\textsc{Score}(\Phi_n, \textsc{Labels}(\Phi_n, c))$\label{al2:score}
			\State $\Psi_n =  \textsc{Cluster}(\Phi_n, c^\text{max}_n)$\label{al2:cluster} \hfill\Comment$\Psi_n= \bigcup_{1 \leq i \leq c^\text{max}_n} \Psi_n^i$
			\State $\Psi = \Psi \cup \Psi_n$ \hfill\Comment collect cluster vectors
			\EndFor
			\State \Return $\Psi$ \label{al2:end}
			\EndFunction
		\end{algorithmic}
	\end{small}
	%			\vspace{-2mm}
\end{algorithm}

Algorithm~\ref{alg:quantisation} shows the high-level process underpinning \approach\ for quantising the vector of neuron activation values and 
extracting clusters for the most important neurons. 
Given as inputs the training set $\mathcal{T} \subseteq \mathcal{X}$, the set of possible clusters $C \subset \mathbb{N_+}$ and the set of important neurons $\mathcal{D}_{m}$
%, identified during neuron-importance analysis 
(cf. Section~\ref{ssec:neuronAnalysis}), 
\approach\ produces for each neuron $n \in \mathcal{D}_m$ the vector of activation values $\Phi_n$ for all training inputs $t \in \mathcal{T}$ (line~\ref{al2:PhiN}). 
Then, through an iterative cluster analysis strategy using the \textit{Silhouette index}~\cite{rousseeuw1987silhouettes}, we find the optimal clustering strategy for each important neuron's activation values (line~\ref{al2:score}). 
Next, we establish the clusters such that $\Psi_n= \bigcup_{1 \leq i \leq c^\text{max}_n} \Psi_n^i$, where $\Psi_n^i$ is the vector containing the activation values for th $i$-th cluster (line~\ref{al2:cluster}). 
We stop when all important neurons have been analysed.

%Score ~1: the value is well matched to the assigned cluster
%Score ~0: the value is borderline matched between two clusters
%Score ~-1: the value may have been assigned to the wrong cluster

Our approach is generic and can support different clustering algorithms, including density-based, grid-based and hierarchical clustering~\cite{kaufman2009finding}. 
We emphasise, however, the importance of using an iterative strategy that enables to determine the optimum number of clusters. This is an important step that defines the granularity of our importance-driven test adequacy criterion (cf. Section~\ref{ssec:relevanceCoverage}). Investigating the applicability and effectiveness of other clustering algorithms and clustering validity criteria is left for future work.

\subsection{Importance-Driven Coverage \label{ssec:relevanceCoverage}}
Given an input set $Y$, we can measure the degree to which it \textit{covers} the clusters of important neurons, termed \textit{Importance-Driven Coverage (IDC)}.  
Since important neurons are core contributors in decision-making (cf. Section~\ref{ssec:neuronAnalysis}), it is significant to establish that inputs triggering combinations of activation value clusters of those neurons (cf. Section~\ref{ssec:neuronQuantisation}) have been covered adequately. 
Doing this, enables to test the most influential neurons, thus increasing our confidence in the correct operation of the DL system and reducing the risk for wrong decisions. 
%and increase our confidence in the correct operation of the DL system. 
The vector of important neurons cluster combinations (INCC) is given by
%More specifically, the evaluation is based on combining the clusters
%\begin{align*}
%K_n = \{ \textsc{Centre}(\Psi_n^i) | \forall \Psi_n^i \in \Psi_n\}
%\end{align*}
\begin{equation}\label{eq:incc}
INCC =  \prod_{n \in D_m}\{ \textsc{Centroid}(\Psi_n^i) | 
\forall 1 \leq i \leq |\Psi_n|\}
%\forall \Psi_n^i \in \Psi_n\}%\\
%&K_{n_1} \times K_{n_2} \times ... \times K_{n_m}
%\{(\textsc{Centre}(\Psi_{n_1}^x), \textsc{Centre}(\Psi_{n_2}^y), ..., \textsc{Centre}(\Psi_{n_m}^z) ) \\
%&\quad\;| \Psi_{n_1}^x \in \Psi_{n_1}, V_{n_2}^y \in \Psi_{n_2}, ..., \Psi_{n_m}^z \in \Psi_{n_m}\}
\end{equation}

\noindent where the function $\textsc{Centroid}(\Psi_n^i)$ measures the ``centre of mass'' of the $i$-th cluster for the $n$-th important neuron.

We define \textit{Importance-Driven Coverage} to be the ratio of INCC covered by all $y \in Y$ over the size of the INCC set. Compared to all other elements in INCC, the $j$-th INNC element is covered if there exists an input $y$ for which the Euclidean distance between the activation values of all important neurons $n \in D_m$ and the corresponding neuron's clusters centroids in $j$ is minimised. Formally
%We define as \textit{Importance-Driven Coverage} to be the ratio of INCC covered by all $t \in \mathcal{T}$. An element in INNC is covered 
\begin{equation}\label{eq:idc}
%IDC(X) \;=\; \frac{\{INCC(i) | \exists x \in X : \forall n \in D_m, V_n^i \in V_n \bullet \min d(\phi(x,n), \textsc{Centre}(V_n^i))\}}
%{\prod_{n \in D_m}c^\text{max}_n}
IDC(Y) = \frac{|\{INCC(j) | 
	\exists y \in Y : 
	\forall V_n^i \in INCC(j) \bullet \min d(\phi(y,n),V_n^i\}|}
{|INCC|}
\end{equation}

Following from (\ref{eq:idc}), a test input is always mapped to an element of the semantic feature set given by INCC (\ref{eq:incc}). IDC increases only if the mapped INCC element tests a new semantic feature set not already covered by existing test suite inputs; otherwise, the score remains the same. We provide a proof of IDC soundness on DeepImportance webpage (https://deepimportance.github.io).

Achieving a high IDC score entails a systematically diverse input set that exercises many combinations of important neurons clusters. 
The covered combinations do not include only those exercised during training, whose activation values have been used for establishing the important neurons, but also new and diverse combinations.
These new combinations could represent edge-case behaviours for the DL system.
%Although the clusters for the important neurons have been established during training,
The higher the IDC score, the more INCC combinations have been triggered. 
Consequently, the more confidence we should have in the DL system's operation.

Another important characteristic of IDC is the \textit{layer-wise} estimation of coverage. By exploiting the combinations of important neurons clusters given by (\ref{eq:incc}), IDC measures how well multiple inputs with semantically different features can trigger those combinations. 
As such, IDC is significantly different to research which focuses on counting  how many neurons have at least once been the most active neurons on a given layer~\cite{MJX18,pei2017deepxplore}.

The granularity with which IDC is specified depends on the number of important neurons $m$ (cf. Algorithm~\ref{alg:analysis}). 
Clearly, setting $m$ to the number of neurons within any layer results in an unmanageable INCC number. 
For instance,  assuming each of the 84 neurons of the penultimate layer of LeNet-5~\cite{lecun1998mnist} produces two clusters (cf. Algorithm~\ref{al2:cluster}), the number of combinations given by (\ref{eq:incc}) is $INCC = 1.9E\!+\!25$. 
Since $m$ is the only $IDC$ hyper-parameter, that  affects the combinations of important neurons clusters (\ref{eq:incc}), it enables software engineers to experiment with different testing strategies by specifying how coarse- or fine-grained the analysis should be. 
In safety-critical systems, for instance, we might opt for a fine-grained IDC coverage, hence a large $m$, aiming to cover as many combinations as possible.
We show in our experimental evaluation that the higher the number of $m$, the higher the number of combinations and the more testing budget is required to increase the IDC score (cf. Section~\ref{sec:evaluation}).
Investigating training-informed ways for the automatic identification of the number of important neurons is part of our future work.

%\sg{TBC: A paragraph or two with details on when/how it can be used and its usefulness 
%\\- layerwise
%\\- difference to top-k neuron coverage
%\\- incorrect corner-case behaviours/edge-case
%} 

%\subsection{Coverage-Driven Input Synthesis \label{ssec:inputSynthesis}}% !TEX root = ../main.tex

% !TEX root = ../main.tex

\vspace*{3mm}\noindent
\section{Implementation}
\label{sec:implementation}

To ease the evaluation and adoption of \approach\ and the Importance-Driven Coverage from Section~\ref{sec:approach}, we have implemented a prototype tool on top of the open-source machine learning framework Keras (v2.2.2)~\cite{chollet2015keras} with Tensorflow (v1.10.1) backend~\cite{tensorflow}. 

The open-source \approach\ source code, the full experimental results summarised in the following section, additional information about \approach\ and the case studies used for its evaluation are available at \url{https://deepimportance.github.io}. 
\noindent% !TEX root = ../main.tex
\vspace*{4mm}\noindent
\section{Evaluation}
\label{sec:evaluation}
\noindent

\subsection{Research Questions}\label{ssec:rqs}
Our experimental evaluation answers the research questions below.

\vspace*{2mm}\noindent
\textbf{RQ1 (Importance): }
% Can our approach select relevant neurons that are sensitive to the most important features of the input?
\textbf{Can neuron-importance analysis identify the most important neurons?}
We used this research question to establish if the importance-based algorithm underpinning \approach\ for the identification of important neurons comfortably outperforms a strategy that selects such neurons randomly. 

\vspace*{2.5mm}\noindent
\textbf{RQ2 (Diversity): }
\textbf{Can \approach\ inform the selection of a diverse test set?}
%\textbf{Can \approach\ result in a semantically diverse test set?}
We investigate whether software engineers can employ the \textit{Importance-Driven Coverage} to generate a diverse test set that comprises semantically different test inputs.
%\approach\ can help testers to form a test set consisting of semantically diverse inputs.

\vspace*{2.5mm}\noindent
\textbf{RQ3 (Effectiveness):}
\textbf{How effective is \approach\ in identifying misbehaviours in DL systems?}
With this research question, we examine the effectiveness of \approach\ to detect adversarial inputs carefully crafted by state-of-the-art adversarial generation techniques~\cite{goodfellow2015explaining,carlini2017adversarial,kurakin2016adversarial,papernot2016limitations}. 
These adversarial inputs should be semantically different than those encountered before, thus increasing the Importance-Driven Coverage metric.
%of \approach.

%For \approach\ to be effective, carefully crafted generated by state-of-the-art adversarial generation techniques~\cite{goodfellow2015explaining,carlini2017adversarial,kurakin2016adversarial,papernot2016limitations} should correspond to semantically different inputs that increase the Importance-Driven Coverage metric of \approach.

\vspace*{2.5mm}\noindent
\textbf{RQ4 (Correlation):}
\textbf{How is \approach\ correlated with existing coverage criteria for DL systems?}
We analyse the relationship in behaviour between \approach\ and state-of-the-art coverage criteria for DL systems including neuron coverage~\cite{pei2017deepxplore}, k-multisection neuron coverage~\cite{MJX18} and surprise adequacy~\cite{Kim2019aa}.

%how the extent to which \approach\ shows similar behaviour with other testing criteria, e.g.,  neuron coverage~\cite{pei2017deepxplore} and k-multisection neuron coverage~\cite{MJX18}. 

%\ \\ \vspace*{-2.5mm} \noindent
%\newpage
\noindent
\textbf{RQ5 (Layer Sensitivity):}
\textbf{How is the behaviour of \approach\ affected by the selection of specific neuron layers?}
Given the layer-wise capability of \approach, we investigate whether performing the analysis on shallow or deeper layers has any impact on the Importance-Driven Coverage metric.

%\vspace*{1mm}\noindent
%\textbf{RQ6 (Robustness):}
%\textbf{Can \approach\ guide retraining of DL systems to improve their accuracy against adversarial examples and synthetic test inputs?}
%A few words here

% \vspace*{1mm}\noindent	
% \textbf{RQ5 (Synthesis Cost):}
% \textbf{What are the overheads for synthesizing new inputs that increase the coverage of \approach?}
% With this research question, we investigate the overheads incurred for synthesizing new inputs that enable to satisfy a given \approach-specific objective.

\begin{table}[b]
	\caption{Datasets and DL Systems used in our experiments.}
%		\vspace*{-2mm}
	\label{table:datasets}
	%	\begin{tabular}{@{}clrl@{}}
	\begin{tabular}{p{2.49cm}p{2.25cm}>{\raggedleft\arraybackslash$}p{1.3cm}<{$}>{\raggedleft\arraybackslash$}p{1.3cm}<{$}}
		\toprule
		\multicolumn{1}{l}{\textbf{Dataset}} & \textbf{DL System} &$\!\!\!$\textbf{\# Params} &$\!\!\!\!\!$\textbf{Performance} \\ \midrule
		\multirow{3}{*}{MNIST~\cite{lecun1998mnist}} & LeNet-1 & 7206 & 98.33\% \\ \cmidrule(l){2-4} 
		& LeNet-4 & 69362 & 98.59\% \\ \cmidrule(l){2-4} 
		& LeNet-5 & 107786 & 98.96\% \\ \midrule
		CIFAR-10~\cite{cifar_model} & \begin{tabular}[c]{@{}l@{}}A 20 layer ConvNet \\ with max-pooling \\and dropout layers. \end{tabular} & 952234 & 77.68\% \\ \midrule 
		Udacity self-driving car challenge\cite{udacity}
		& Dave-2 archite- cture from Nvidia & 2116983 &0.096$\qquad$(MSE)\\
		\bottomrule
		\multicolumn{4}{l}{
			\begin{minipage}{\linewidth}\small
				The column `Params' shows the number of trainable parameters for each DL model. The column `Performance' shows the accuracy (for for MNIST and CIFAR-10 datasets) and mean squared error (for the Udacity dataset).
			\end{minipage}
		}\\
	\end{tabular}
\end{table}

\subsection{Experimental Setup}\label{ssec:experimentalSetup}

\textbf{Datasets and DL Systems.}
Table~\ref{table:datasets} shows the datasets and DL systems used in our experimental evaluation. 
We evaluate \approach\ on three popular publicly-available datasets.  
MNIST~\cite{lecun1998mnist} is a  handwritten digit dataset with 60,000 
training inputs  and 10,000 testing  inputs; each input is a 28x28 pixel image with a class label from 0 to 9. CIFAR-10~\cite{cifar_model} is an image dataset with 50,000 training inputs  and 10,000 testing inputs; each input is a 32x32 image in ten different classes (e.g., dog, bird, car). 
The Udacity self-driving car challenge dataset~\cite{udacity} comprises images captured by a camera mounted behind the windshield of a moving car and supported by the steering wheel angle applied by the human driver for each image. 
Since this is the ground truth, the aim for a DL system is to predict the steering wheel angle; hence, the DL system's accuracy is measured using Mean Squared Error (MSE) between ground truth and predicted steering angles. 
The Udacity dataset has 101,396 training and 5,614 testing inputs. 
%\ \\
%\sg{Last, we consider Udacity self-driving car challenge dataset \cite{udacity} that consists of images captured by a camera mounted behind the windshield of a moving car and the simultaneous steering wheel angle applied by the human driver for each image. The dataset has 101,396 training and 5,614 testing inputs.} 

To enable a systematic and comprehensive assessment of \approach, we chose DL systems used in related research~\cite{pei2017deepxplore,wicker2018feature,MJX18,Kim2019aa} with different architecture (i.e., number of layers and layer types - fully-connected, convolutional, dropout, max- pooling), complexity (i.e., number of trainable parameters)  and performance. 
For MNIST, we study three DL systems from the Le-Net family~\cite{lecun1998mnist}, i.e.,  LeNet-1, LeNet-4 and LeNet-5, trained to achieve over 98\% accuracy on the provided test set (cf. Table~\ref{table:datasets}).
For CIFAR-10, we employ the prototype model in~\cite{cifar_model} which is a 20 layer convolutional neural network (CNN) trained to achieve 77.68\% accuracy. 
For the Udacity self-driving car challenge, we used the pre-trained Dave-2~\cite{bojarski2016end} self-driving car DL system from Nvidia. 
Dave-2 comprises nine layers including five convolutional layers and its MSE is 0.096. 
All experiments were run on an Ubuntu server with 16 GB memory and Intel Xeon E5-2698 2.20GHz.

\noindent
\textbf{Coverage Criteria Configurations.} 
We facilitate a thorough and unbiased evaluation of \approach\ by comparing it against state-of-the-art coverage criteria for DL systems. 
To this end, we used DeepXplore's~\cite{pei2017deepxplore} neuron coverage (NC); 
DeepGauge's~\cite{MJX18} k-multisection neuron coverage (KMNC), neuron boundary coverage (NBC), strong neuron activation coverage (SNAC) and top-k neuron coverage (TKNC); and
Surprise's Adequacy~\cite{Kim2019aa} distance-based (DSC) and likelihood-based surprise coverage (LSC). 
For each criterion, we use the hyper-parameters recommended in its original research. In particular, we set neuron activation threshold to 0.75 in \textit{NC}, and $k=3$ and $k=1000$ in TKNC and KMNC, respectively.
For NBC and SNAC we set as lower (upper) bound the minimum (maximum) activation value encountered in the training set. 
The upper bound for DSC and LSC is fixed to 2 and 2000, respectively, and the number of buckets is set to 1000. 
Concerning \approach, unless otherwise stated (e.g., RQ5), we always consider the penultimate layer as the subject layer and the number of important neurons $m \in \{6, 8, 10, 12\}$. 
\
When running the experiments, we set an upper bound of execution time to three hours. If a criterion exceeds this threshold, we terminate its execution and report that no results have been generated. 
We facilitate the replication of our findings by making available the implementation of all those metrics on the project webpage.

\vspace*{2mm}\noindent
\textbf{Synthetic Inputs and Adversarial Examples.}
We use both synthetic inputs and adversarial examples to evaluate \approach.
Synthetic inputs are obtained by applying small perturbations on the original inputs through Gaussian-like injected white noise~\cite{an1996effects,arpit2017closer}. 
Adversarial examples are carefully crafted perturbations to inputs, which albeit being imperceptible to the human, lead a DL system to make an incorrect decision~\cite{goodfellow2015explaining}.
Adversarial examples are typically used to assess the robustness of DL systems. 
We employ four widely studied attack strategies to evaluate \approach\: Fast Gradient Sign Method (FGSM)~\cite{goodfellow2015explaining}, Basic Iterative Method (BIM)~\cite{kurakin2016adversarial}, Jacobian-based Saliency Map Attack (JSMA)~\cite{papernot2016limitations}, and Carlini\&Wagner (C\&W)~\cite{carlini2017towards}. Our implementation of these strategies is based on Cleverhans~\cite{papernot2016cleverhans}, a Python library for benchmarking DL systems against adversarial examples.

%In adversarial inputs, even though the perturbations are imperceptible DNN decisions are changed~\cite{goodfellow2015explaining}. Adversarial examples are used in assessing the robustness of DNNs, since they  are more effective in revealing robustness issues in DNNs than the test inputs provided by the original datasets. We employ four widely studied attack strategies to evaluate \approach\: Fast Gradient Sign Method (FGSM)~\cite{goodfellow2015explaining}, Basic Iterative Method (BIM)~\cite{kurakin2016adversarial}, Jacobian-based Saliency Map Attack (JSMA)~\cite{papernot2016limitations}, and Carlini\&Wagner (C\&W)~\cite{carlini2017towards}. Our implementation of these strategies is based on Cleverhans~\cite{papernot2016cleverhans}, a Python library to benchmark DL systems against adversarial examples.

\subsection{Results and Discussion}\label{ssec:results}

\noindent\textbf{RQ1 (Importance).} 
Since identifying the most important neurons within a subject layer is a key principle of \approach, we assess if the neurons identified during neuron-importance analysis (cf. Algorithm~\ref{alg:analysis}) have indeed a significant role in decision-making. 
To answer this research question, we employ \approach\ to find the $m=6$ and $m=8$ most important neurons for the MNIST and Cifar-10, and Udacity datasets, respectively. 
We select an equivalent number of neurons using a random-selection strategy. 
Next, we employed the approach from~\cite{bach2015pixel}, used in the \textit{explainable AI} area to highlight input parts responsible for a decision, and chose inputs (pixels) whose score is above the 90th percentile (i.e., among the top 10\%). 
We then perturbed those pixels, setting their value to zero if their score is above a predefined threshold of 0.5 (i.e., they are relevant) and 
%setting their value 
to one otherwise. 
We limit the magnitude of perturbation to at most 10\% of the total number of pixels, aiming to keep the perturbed input close to the original.
% we also define a threshold such that the maximum number of pixels that can be flipped is not greater than 10\% of the total number of pixels, for the total perturbation amount that is applied to an input so that the perturbed input is still a valid image and it resembles the original image. 
Finally, we measured the L2 (Euclidean) distance between the activation values of the original input and the perturbed input both for \approach\ and random; 
a higher distance signifies a more significant change.
%use the DL systems described in Section~\ref{ssec:experimentalSetup} and 

\begin{figure}[t]
	\centering
	\vspace*{2mm}
	
	\includegraphics[clip,scale=0.41]{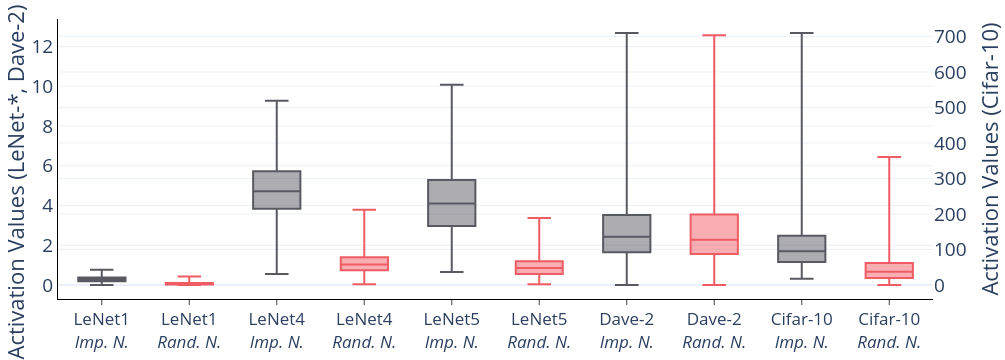}
	
%	\vspace*{-3mm}
	\caption{Boxplots comparing activation values distance of important and randomly-selected neurons between original inputs and those with their most relevant pixels perturbed.}
	\label{fig:boxplot_val}
\end{figure}

Figure~\ref{fig:boxplot_val} and Table~\ref{table:validation_rand_rel} show boxplots and the average delta (standard deviation in parenthesis) of activation values for the entire testing set (i.e., for all classes) of each dataset, respectively. 
The reported results are over five independent runs, thus mitigating the risk that they have been obtained by chance. 
Clearly, the activation values distance for neurons selected by \approach\ is higher than the equivalent distance for randomly-selected neurons. 
%Clearly, there is a significant difference in the activation values distance for \approach-selected compared to randomly-selected neurons. 
The difference becomes more evident in LeNet-4 and LeNet-5 that have 120 and 84 neurons in the penultimate layer, respectively, with the distance using \approach\ exceeding 4.18, whereas the distance using random is between 0.93 and 1.09. 
Similar observations hold for Cifar-10 (128 feature maps), while the difference is less clear for LeNet-1 (12 feature maps). 
These observations also provide a useful indication for the number of important neurons $m$ with regards to the total number of neurons in the subject layer.
%difference becomes more 

\textbf{
	We conclude that \approach\ can detect the most important neurons of a DL system and those neurons are more sensitive to changes in relevant pixels of a given input.
}

\begin{table}[t]
	\caption{Average (std dev.)  L2 distance of activation values for neurons selected randomly and using \approach\ on MNIST (LeNet-1|4|5), Cifar-10 and Udacity (Dave-2).}
	%\caption{Validation of \approach\ activation values for the most improtant neurons (distance is measured in euclidean distance.)}
	\label{table:validation_rand_rel}
	\vspace*{-4mm}
	\scalebox{0.8}{
		\begin{tabular}{llllll}
			\toprule
			$\!\!\!\!$\textbf{Strategy}& 
			$\!\!\!$\textbf{LeNet-1} &$\!\!$\textbf{LeNet-4} &$\!\!$\textbf{LeNet-5} &$\!\!$\textbf{Cifar-10} &$\!\!\!$\textbf{Dave-2}\\ \midrule
			$\!\!\!\!$\textit{Random} &  
			$\!\!\!$0.07($\pm$0.05) &$\!\!$1.09($\pm$0.49)  &$\!\!$0.93 ($\pm0.51$) &$\!\!\!$47.22($\pm$42.8) &$\!\!\!$1.16($\pm$0.75)\\
			$\!\!\!\!$\textit{\approach} 
			&$\!\!\!$0.28($\pm$0.13) &$\!\!$4.79($\pm$1.35) &$\!\!$4.18($\pm1.61$) &$\!\!$112.03($\pm$70.3) &$\!\!\!$2.83($\pm$1.92)\\ \bottomrule
		\end{tabular}
	}
\end{table}

\vspace*{3mm}
\noindent\textbf{RQ2 (Diversity).} 
%Evaluating the ability of a DL system to learn semantically diverse
A useful coverage criterion for DL systems  entails the ability to assign higher coverage for test sets that comprise \textit{semantically diverse} test inputs~\cite{zhang2019machine}. 
This is a significant asset for evaluating the ability of a DL system to learn \textit{semantically meaningful features} for the decision-making task rather than \textit{memoising} or learning irrelevant features (i.e., learn to make decisions by exploiting unintended similarity patterns in the test set)~\cite{amodei2016concrete}. 

\begin{figure}[t]
	\centering
	\includegraphics[scale=0.36]{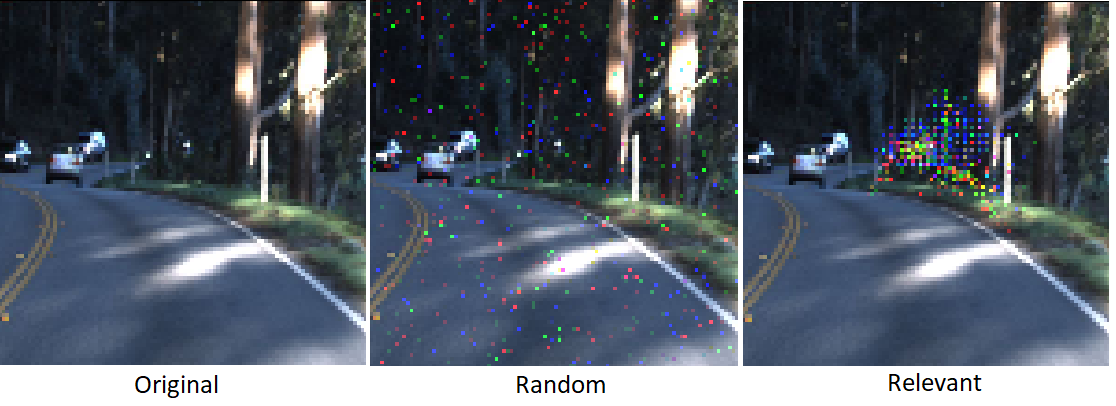}
	
	 \vspace*{-2mm}
	\caption{Example image from the Udacity dataset showing the original input (left), an input from the $U_S$ set with Gaussian noise in random pixels (centre), and an input from the $U_{DI}$ set with Gaussian noise to relevant pixels (right).}
	%    	. The leftmost image is the original input, in the centered image white noise added to random pixels and in the rightmost image whit enoise added to relevant pixels.}
	\label{fig:dave_sample}
		\vspace*{-6mm}
\end{figure}

To answer this research question, we measured the IDC metric~(\ref{eq:idc}) given the original test set $U_O$ of each dataset and corresponding DL systems (cf. Section~\ref{ssec:experimentalSetup}) for multiple values of important neurons $m \in \{6, 8, 10, 12\}$. 
Then, for each test set we created two `perturbed' versions.
The former is a \textit{semantically diverse} set $U_{DI}$ that consists of inputs whose top 2\% pixels (identified similarly as in RQ1), are perturbed by applying small perturbations to the original inputs through Gaussian white noise~\cite{an1996effects,arpit2017closer}. The number of perturbed pixels are 15 for MNIST, 20 for Cifar-10 and 200 for Udacity. 
The other is a \textit{numerically diverse} set $U_S$ that consists of synthetic inputs generated also by injecting Gaussian white noise to an equivalent number of randomly-selected pixels of original inputs. 
For example, Fig.~\ref{fig:dave_sample} shows an original input from the Udacity dataset (left), the perturbed input from the $U_S$ set (centre) and the perturbed input within the $U_{DI}$ set (right). 
The modified pixels in the image on the right are the top 2\% pixels that lead the car to steer the wheel to the left (ground truth). 
We add both of these perturbed test sets to the original set and obtain the test sets $U_{O+DI}$ and $U_{O+S}$ and measured their IDC metric.

Table \ref{tab:relevance_validation} (top rows with IDC prefix) shows the average IDC value for different DL systems and number of important neurons $m$. 
As before, we reduce randomisation bias by reporting results over five independent runs. 
For all datasets and DL systems, the IDC value for the semantically diverse set
(column $U_{O+DI}$) is always higher than that for the numerically diverse set ($U_{O+S}$ column). 
In fact, the difference becomes more clear for the more complex DL systems, e.g., LeNet-4 (+1\% on average) and Dave-2 (+1.5\% on average). 
This behaviour is also reinforced by a corresponding reduction in accuracy.
In particular, in all instances the prediction confidence for the $U_{O+S}$ set is always higher than that of the $U_{O+DI}$ set.
These observations signify that IDC is more sensitive to input features that are important to the decision-making task instead of randomly-selected 
features.

\begin{table*}[ht!]
	\caption{Average (std dev) coverage results for Importance-Driven Coverage criterion ($m\!\in\!\{6,\!8,\!10,\!12\}$) and other %state-of-the-art 
		coverage criteria for MNIST, Cifar-10 and Udacity datasets; the highest value between $U_S$ and $U_{DI}$ is boldfaced (T/O: timeout, N/A: not applicable).
	}
	\vspace*{-4mm}
	\label{tab:relevance_validation}
	\setlength{\tabcolsep}{1.5pt}
	\renewcommand{\arraystretch}{1.5}
	\scalebox{0.7}{
		\begin{tabular}{@{}| l |
				>{\columncolor[HTML]{EFEFEF}}c 
				>{\columncolor[HTML]{EFEFEF}}c 
				>{\columncolor[HTML]{EFEFEF}}c | ccc |
				>{\columncolor[HTML]{EFEFEF}}c 
				>{\columncolor[HTML]{EFEFEF}}c 
				>{\columncolor[HTML]{EFEFEF}}c | ccc |
				>{\columncolor[HTML]{EFEFEF}}c 
				>{\columncolor[HTML]{EFEFEF}}c 
				>{\columncolor[HTML]{EFEFEF}}c | @{}}
			\toprule
			\multicolumn{1}{|c|}{\textbf{}} & \multicolumn{3}{c|}{\cellcolor[HTML]{EFEFEF}\textbf{LeNet-1 (MNIST)}} & \multicolumn{3}{c|}{\textbf{LeNet-4 (MNIST)}} & \multicolumn{3}{c|}{\cellcolor[HTML]{EFEFEF}\textbf{LeNet-5 (MNIST)}} & \multicolumn{3}{c|}{\textbf{Cifar-10}} & \multicolumn{3}{c|}{\cellcolor[HTML]{EFEFEF}\textbf{Dave-2 (Udacity)}} \\ \midrule
			& $U_O$ & $U_{O+S}$ & $U_{O+DI}$ 
			& $U_O$ & $U_{O+S}$ & $U_{O+DI}$ 
			& $U_O$ & $U_{O+S}$ & $U_{O+DI}$ 
			& $U_O$ & $U_{O+S}$ & $U_{O+DI}$ 
			& $U_O$ & $U_{O+S}$ & $U_{O+DI}$ \\ \midrule
			\textbf{$\!$IDC$_6$} 
			&$\!$34.6\%($\pm2.2$) & 38.0\%($\pm2.5$) & \textbf{38.8\%($\pm2.4$}) 
			& 58.8\%($\pm2.7$) & 64.2\%($\pm2.7$) & \textbf{65.8\%($\pm2.7$)} 
			& 47.0\%($\pm3.0$) & 51.1\%($\pm2.9$) & \textbf{52.1\%($\pm2.8$)} 
			& 29.4\%($\pm1.3$) & 37.9\%($\pm1.5$) & \textbf{39.0\%($\pm1.4$)} 
			& 19.1\%($\pm0.7$) & 26.0\%($\pm1.1$) & \textbf{28.5\%($\pm1.1$)} \\
			\textbf{$\!$IDC$_8$} 
			&$\!$14.1\%($\pm0.9$) & 16.5\%($\pm1.1$) & \textbf{17.5\%($\pm1.2$)} 
			& 26.9\%($\pm1.1$) & 31.8\%($\pm1.4$) & \textbf{32.9\%($\pm1.5$)} 
			& 28.0\%($\pm1.2$) & 33.7\%($\pm1.7$) & \textbf{34.5\%($\pm1.7$)} 
			& 11.1\%($\pm0.6$) & 15.5\%($\pm0.8$) & \textbf{16.6\%($\pm0.9$)} 
			& 7.8\%($\pm0.3$) & 10.0\%($\pm0.5$)  & \textbf{11.6\%($\pm0.2$)} \\
			\textbf{$\!$IDC$_{10}$} 
			&$\!$5.5\%($\pm0.3$) & 6.6\%($\pm0.4$) & \textbf{7.0\%($\pm0.4$}) 
			& 13.5\%($\pm0.8$) & 16.9\%($\pm1.0$) & \textbf{18.1\%($\pm1.0$)} 
			&  9.5\%($\pm0.4$) & 12.2\%($\pm0.5$) & \textbf{13.1\%($\pm0.6$)} 
			& 5.1\%($\pm0.3$) & 7.4\%($\pm0.4$) & \textbf{8.1\%($\pm0.4$)} 
			& 3.3\%($\pm0$) & 4.2\%($\pm0.1$) & \textbf{4.9\%($\pm0.2$)} \\
			\textbf{$\!$IDC$_{12}$} 
			&$\!$2.1\%($\pm0.1$) & 2.6\%($\pm0.1$) & \textbf{2.9\%($\pm0.2$)} 
			& 4.5\%($\pm0.2$) & 6.3\%($\pm0.3$) & \textbf{6.9\%($\pm0.4$)} 
			& 4.4\%($\pm0.2$) & 6.2\%($\pm0.4$) & \textbf{6.8\%($\pm0.4$)} 
			& 2.0\%($\pm0.2$) & 3.2\%($\pm0.3$) & \textbf{3.7\%($\pm0.3$)} 
			& 1.4\%($\pm0$) & 2.0\%($\pm0$) & \textbf{2.3\%($\pm0$)} \\ \midrule
			\textbf{$\!$NC} 
			&$\!$17.3\%($\pm0.5$) & \textbf{20.7\%($\pm0.3$)} & 20.5\%($\pm0.3$) 
			& 37.9\%($\pm0.7$) & \textbf{44.1\%($\pm0.8$)} & 43.4\%($\pm0.8$) 
			& 44.2\%($\pm0.9$) &  \textbf{51.7\%($\pm0.9$)} &  50.6\%($\pm0.9$) 
			& 20.2\%($\pm0.3$) & \textbf{35.2\%($\pm0.2$)} & 34.5\%($\pm0.2$) 
			& 51.6\%($\pm1.8$) & \textbf{66.7\%($\pm0.2$)} & 64.7\%($\pm0.3$) \\
			\textbf{$\!$KMNC} 
			&$\!$35.1\%($\pm0.3$) & \textbf{51.3\%($\pm0.4$)} & 48.2\%($\pm0.4$) 
			& 34.9\%($\pm0.2$) & \textbf{54.4\%($\pm0.3$)} &  50.8\%($\pm0.3$) 
			& 32.5\%($\pm0.2$) & \textbf{52.0\%($\pm0.3$)} & 48.4\%($\pm0.2$) 
			& 36.8\%($\pm0$) & \textbf{43.5\%($\pm0$)} & 41.5\%($\pm0$) 
			& 30.2\%($\pm0$) & \textbf{50.9\%($\pm0.1$)} & 46.6\%($\pm0.1$) \\
			\textbf{$\!$NBC} 
			&$\!$16.7\%($\pm1.3$) & \textbf{22.5\%($\pm1.4$)} & 21.5\%($\pm1.3$) 
			& 9.3\%($\pm0.6$) & \textbf{12.3\%($\pm0.6$)} & 11.7\%($\pm0.6$) 
			& 9.3\%($\pm0.5$) & \textbf{12.1\%($\pm0.6$)} & 11.5\%($\pm0.5$) 
			& 9.4\%($\pm0$) & \textbf{9.5\%($\pm0$)} & \textbf{9.5\%($\pm0$) }
			& 0.8\%($\pm0.1$) & \textbf{24.7\%($\pm0.5$)} & 21.4\%($\pm0.7$) \\
			\textbf{$\!$SNAC} 
			&$\!$10.9\%($\pm0.6$) & \textbf{14.6\%($\pm0.6$)} & 13.8\%($\pm0.6$) 
			& 12.0\%($\pm0.5$) & \textbf{15.0\%($\pm0.6$)} & 14.3\%($\pm0.6$) 
			&  14.4\%($\pm0.5$) & \textbf{18.1\%($\pm0.6$) }& 17.3\%($\pm0.6$) 
			& 8.8\%($\pm0$) & \textbf{8.9\%($\pm0$)} & \textbf{8.9\%($\pm0$)} 
			& 1.5\%($\pm0.2$) & \textbf{46.9\%($\pm0.9$)} & 41.0\%($\pm1.4$) \\
			\textbf{$\!$TKNC} 
			&$\!$100.0\%($\pm0.0$) & \textbf{100.0\%($\pm0.0$)} & \textbf{100.0\%($\pm0.0$)}& 91.3\%($\pm0.0$) & \textbf{91.7\%($\pm0.2$) }& 91.6\%($\pm0.2$) & 88.8\%($\pm0.0$) & \textbf{89.2\%($\pm0.0$)} & 89.1\%($\pm0.1$) & 15.2\%($\pm0.0$) & \textbf{17.0\%($\pm0.1$)} & 16.6\%($\pm0.1$) & 40.8\%($\pm0.1$) & \textbf{52.1\%($\pm0.2$)} & 50.0\%($\pm0.2$) \\
			\textbf{$\!$DSC} 
			&$\!$86.3\%($\pm0.0$) & 91.7\%($\pm0.3$) & \textbf{92.3\%($\pm$0.5)} & 60.2\%($\pm$0.3) & \textbf{66.8\%($\pm$0.1)} & 66.7\%($\pm$0.2) & 54.9\%($\pm0.0$) & 60.9\%($\pm$0.2) & \textbf{61.4\%($\pm$0.2)} & TO & TO & TO & N/A & N/A & N/A \\
			\textbf{$\!$LSC} 
			&$\!$2.8\%($\pm0.1$) & \textbf{3.3\%($\pm0.1$)} & 3.2\% ($\pm0.1$) & 14.6\%($\pm0.1$) & \textbf{16.7\%($\pm$0.1)} & 16.5\%($\pm$0.2) & \% 13.8($\pm0.0$) & 16.5\%($\pm$0.1) & \textbf{16.8\%($\pm$0.2)} & TO & TO & TO  & 4.2\%($\pm$0.1) & 4.6\%($\pm$0.1) &  \textbf{4.7\%($\pm$0.1) }\\ \bottomrule
	\end{tabular}}
\end{table*}

Another interesting observation from Table~\ref{tab:relevance_validation} is that due to the INCC number (\ref{eq:incc}), the IDC value becomes lower as the number of important neurons $m$ increases. 
Considering LeNet-4, for instance, IDC decreases from 65.8\% (64.2\%) to 18.1\%(16.9\%) for $U_{O+DI}$($U_{O+S}$) when $m=6$ and $m=10$, respectively. 
For these experiments, the number of clusters of important neurons extracted from Algorithm~\ref{alg:quantisation} is between two and four. 
This is expected since the combinations of important neurons clusters (INCC) increases exponentially as $m$ increases (e.g., $[64,486]$ for $m=6$ and $[4096, 69984]$ for $m=12$).  
Software engineers can use this information to adjust the available budget and effort needed to test their DL systems. 
 
For completeness, we ran similar experiments using state-of-the-art coverage criteria for DL systems (cf. Section~\ref{ssec:experimentalSetup}). 
Table~\ref{tab:relevance_validation} (bottom) shows their coverage results. 
Except from LeNet-1, i.e., the DL system with the smallest complexity, the coverage results for all other DL systems are smaller for the \textit{semantically diverse} set $U_{DI}$ compared to 
%they are indifferent between the \textit{semantically-diverse} set $U_{DI}$ and 
the \textit{numerically diverse} set $U_{S}$.  
In contrast to IDC, which is sensitive to perturbations to relevant  input features, these criteria are also sensitive to perturbations to random input features.\hspace*{-1mm}

\textbf{
	We conclude that \approach\ with its IDC coverage criterion can support software engineers to create a diverse test set that comprises semantically different test inputs.
}

\vspace*{2mm}
\noindent\textbf{RQ3 (Effectiveness).} 
Building on research in traditional software testing, effective coverage criteria for DL systems should be capable of identifying misbehaviours (i.e., failing test cases)~\cite{huang2018safety}.
%The ability to identify misbehaviours (i.e., failing test cases) is another core property of practicable coverage criteria for DL systems~\cite{huang2018safety}. 
Coverage criteria satisfying this property have good fault-detection abilities. Thus, they can be used to evaluate the \textit{adequacy} of a test set and provide a quantifiable measurement of confidence in testing~\cite{pezze2008software}.  

To assess the effectiveness of \approach, we compared the IDC values between an unmodified test set $U_O$ and sets enhanced with perturbed inputs using white noise and adversarial inputs carefully crafted using state-of-the-art adversarial generation techniques. 
More specifically, for each dataset and each DL system,  we generated four adversarial test sets using FGSM~\cite{goodfellow2015explaining}, BIM~\cite{kurakin2016adversarial}, JSMA~\cite{papernot2016limitations} and CW~\cite{carlini2017adversarial} and another \textit{numerically diverse} test $U_S$ via Gaussian white noise with standard deviation=0.3 (as in RQ1). 
Unlike adversarial inputs, the set $U_S$ is correctly classified with accuracy 97.4\% on average.
We extended the original set with each of these synthesised test sets and measured their IDC values for the corresponding DL systems. 

Table~\ref{tab:effect} (columns $IDC_6$ and $IDC_8$) shows the average IDC coverage results for $m \in \{6, 8\}$ across the six enhanced test sets of each DL system.  
Compared to the original test set $U_O$, there is a considerable increase in the IDC result for the enhanced test sets for all DL systems. 
As expected, the IDC result for $m=6$ ($IDC_6$) is higher than that for $m=8$ ($IDC_8$) as the number of combinations INCC (\ref{eq:incc}) grows exponentially with the number of important neurons. 
%More specifically, 
The increase is more significant in test sets involving adversarial inputs than that with Gaussian-like noisy inputs ($U_{O+S}$)  .
Consequently, adversarial inputs lead to higher coverage for our IDC criterion, thus signifying the \textit{sensitivity} to adversarial inputs and its fault detection abilities (conforming to  testing criteria in traditional software testing). 

\textbf{%Based on these results, 
	We conclude that IDC is sensitive to adversarial inputs and is effective in detecting misbehaviours in test sets with inputs semantically different than those encountered before.}
	% IDC is sensitive to adversarial inputs and can provide objective insight on the correctness of the DNN when satisfied.}
%internal behavioral difference of benign and adversarial test data.

\begin{table}[t]
	\caption{Effectiveness of coverage metrics. %in detecting adversarial examples
		(`+Y' means adding Y-based adversarial inputs to the original test set $U_O$)}
	\vspace*{-2mm}
	\label{tab:effect}
	\setlength{\tabcolsep}{1.4pt}
	\renewcommand{\arraystretch}{1.5}
	\scalebox{0.9}{
		\begin{tabular}{@{}|c|l|ccccccccc|@{}}
			\toprule
			&  & \textbf{\textit{IDC}$_6$} & \textbf{\textit{IDC}$_8$} & \textbf{NC} & \textbf{KMNC} & \textbf{NBC} & \textbf{SNAC} & \textbf{TKNC} & \textbf{LSA} & \textbf{DSA} \\ \midrule
			\rowcolor[HTML]{EFEFEF} 
			\cellcolor[HTML]{EFEFEF} & $U_O$. & 34.6\% & 14.1\% & 23.8\% & 62.7\% & 15.1\% & 18.6\% & 100\% & 2.6\% & 86.2\% \\
			\rowcolor[HTML]{EFEFEF} 
			\cellcolor[HTML]{EFEFEF} & $U_{O+S}$ & 36.3\% & 16.1\% & 23.8\% & 70.8\% & 25.0\% & 18.6\% & 100\% & 4.0\% & 87.6\% \\
			\rowcolor[HTML]{EFEFEF} 
			\cellcolor[HTML]{EFEFEF} & +FGSM & 42.3\% & 20.9\% & 23.8\% & 81.1\% & 46.5\% & 55.8\% & 100\% & 13.7\% & 85.3\% \\
			\rowcolor[HTML]{EFEFEF} 
			\cellcolor[HTML]{EFEFEF} & +BIM & 43.2\% & 20.8\% & 23.8\% & 71.6\% & 45.3\% & 53.4\% & 100\% & 9.6\% & 86.8\% \\
			\rowcolor[HTML]{EFEFEF} 
			\cellcolor[HTML]{EFEFEF} & +JSMA & 41.0\% & 19.0\% & 23.8\% & 80.5\% & 31.3\% & 37.2\% & 100\% & 13.9\% & 86.9\% \\
			\rowcolor[HTML]{EFEFEF}
			\multirow{-6}{*}{\cellcolor[HTML]{EFEFEF}\rotatebox{90}{\textbf{LeNet-1}}} & +CW & 37.9\% & 17.0\% & 23.8\% & 64.9\% & 16.6\% & 19.0\% & 100\% & 5.2\% & 86.2\% \\  \midrule
			& $U_O$. & 58.8\% & 27.0\%  & 63.7\% & 69.2\% & 7.9\% & 12.3\% & 91.3\% & 14.4\% & 61.5\% \\
			& $U_{O+S}$ & 62.0\% & 29.1\% & 64.4\% & 72.6\% & 10.8\% & 12.3\% & 91.3\% & 10.9\% & 67.0\% \\
			& +FGSM & 65.6\% & 33.4\% & 64.4\% & 79.4\% & 38.8 \% & 65.4\% & 93.4\% & 39.3\% & 83.7\% \\
			& +BIM & 66.5\% & 33.4\% & 79.3\% & 74.2\% & 41.0\% & 69.7\% & 92.7\% & 45.1\% & 78.8\% \\
			& +JSMA & 64.7\% & 32.2\% & 63.7\% & 76.8\% & 14.3\% & 20.8\% & 91.3\% & 64.4\% & 88.8\% \\
			\multirow{-6}{*}{\rotatebox{90}{\textbf{LeNet-4}}} & +CW & 62.8\% & 31.0\% & 63.7\% & 70.5\% & 7.9\% & 12.3\% & 91.3\% &  14.4\% & 60.1  \\ \midrule
			\rowcolor[HTML]{EFEFEF} 
			\cellcolor[HTML]{EFEFEF} & $U_O$. & 47.0\% & 28.0\% & 75.3\% & 69.2\% & 7.6\% & 13.8\% & 88.8\% & 13.8\% & 54.9\% \\
			\rowcolor[HTML]{EFEFEF} 
			\cellcolor[HTML]{EFEFEF} & $U_{O+S}$ & 48.1\% & 29.1\% & 75.3\% & 71.5\% & 10.0\% & 13.8\% & 88.8\% & 10.8\% & 57.3\% \\
			\rowcolor[HTML]{EFEFEF} 
			\cellcolor[HTML]{EFEFEF} & +FGSM & 51.6\% & 32.3\% & 75.3\% & 79.6\% & 40.7\% & 71.3\% & 89.1\% & 40.4\% & 83.5\% \\
			\rowcolor[HTML]{EFEFEF} 
			\cellcolor[HTML]{EFEFEF} & +BIM & 51.6\% & 32.3\% & 84.7\% & 76.1\% & 46.2\% & 83.2\% & 89.1\% & 42.4\% & 74.3\% \\
			\rowcolor[HTML]{EFEFEF} 
			\cellcolor[HTML]{EFEFEF} & +JSMA & 49.8\% & 32.3\% & 75.3\% & 74.3\% & 12.8\% & 21.1\% & 89.1\% & 61.0\% & 85.7\% \\
			\rowcolor[HTML]{EFEFEF} 
			\multirow{-6}{*}{\cellcolor[HTML]{EFEFEF}\rotatebox{90}{\textbf{LeNet-5}}} & +CW & 49.8\% & 31.8\% & 75.3\% & 70.6\% & 7.6\% & 13.8\% & 88.8\% &  15.2\% & 61.2\%  \\ \bottomrule
	\end{tabular}}
	
	%\vspace*{-4mm}
\end{table}

\vspace*{2mm} 
\noindent\textbf{RQ4 (Correlation).} 
We report results on how state-of-the-art coverage criteria behave across the six tests sets for MNIST in Table~\ref{tab:effect}. 
Similarly to IDC, most of the criteria, i.e., KMNC, NBC, SNAC, LSA, DSA, experience a similar increase to their coverage results when evaluated using test sets augmented with adversarial inputs (e.g., FGSM, BIM, JSMA, CW).
As such, IDC is consistent with criteria based on input surprise (e.g., LSA, DSA) and aggregation of neuron property values (e.g., KMNC, NC).  
However, while the IDC result for the test set $U_{O+S}$ is always lower than that with adversarial inputs (with the exception of BIM for $IDC_8$ on LeNet-4), there are several instances in which $U_{O+S}$ produces higher results than the adversarial-augmented sets (e.g., KMNC, NBC DSA for LeNet-1). 
This is an interesting finding that requires further investigation.

Another interesting observation is that NC and TKNC are insensitive to either Gaussian-like noisy inputs or adversarial inputs, irrespective of the employed adversarial technique. 
The results for NC are not surprising and conform to results reported in existing research~\cite{MJX18,Kim2019aa}.
Nevertheless, the plateau shown in TKNC is particularly important since it is a layer-wise coverage criterion, like IDC. 
In contrast to IDC, TKNC measures how many neurons have at least once been the most active $k$ neurons on a target (or all) layer. 
Considering these results, IDC is more informative than TKNC. 

%We report results on how state-of-the-art coverage criteria behave across the six tests sets for MNIST in Table~\ref{tab:effect}. 
%Similarly to IDC, most of the criteria, i.e., KMNC, NBC, SNAC, LSA, DSA, experience a similar increase to their coverage results when evaluated using test sets augmented with adversarial inputs (e.g., FGSM, BIM, JSMA, CW).
%This signifies that IDC is consistent with criteria based on input surprise (e.g., LSA, DSA) and aggregation of neuron property values (e.g., KMNC, NC).  

\textbf{In general, we conclude that IDC shows a similar behaviour to state-of-the-art coverage criteria for DL systems; hence, there is a positive correlation between them.}

\begin{table}[t]
	\caption{IDC coverage results for different layers with the best coverage between the $U_O$ and $U_{O+DI}$ sets typeset in bold.}
	\label{tab:layer_sens}
	\vspace*{-2mm}
	\setlength{\tabcolsep}{4.2pt}
	\scalebox{1}{
		\begin{tabular}{@{}ccccccc@{}}
			\toprule
			\textbf{} & \multicolumn{2}{c}{\textbf{LeNet1 ($IDC_4$)}} & \multicolumn{2}{c}{\textbf{LeNet4 ($IDC_4$)}} & \multicolumn{2}{c}{\textbf{LeNet5 ($IDC_6$)}} \\ \midrule
			& $U_O$ & $U_{O+DI}$
			& $U_O$ & $U_{O+DI}$
			& $U_O$ & $U_{O+DI}$\\ \midrule
			Conv1 
				& 35.3\% & \textbf{38.3\%}
				& 33.9\% & \textbf{35.9\%} 
				& 12.5\% & \textbf{15.6\%} \\
			Conv2 
				& 76.2\% & \textbf{80.8\%}
				& 81.6\% & \textbf{84.0\%} 
				& 31.0\% & \textbf{36.6\%} \\
			FC1 & - & - & 86.0\% & \textbf{90.0\%} & 37.0\% & \textbf{44.2\%} \\
			FC2 & - & - & - & - & 35.8\% & \textbf{43.4\%} \\ \bottomrule
			\multicolumn{7}{l}{
			{\small Conv*: Convolutional layer; FC*: Fully-connected layer}}\\
			\multicolumn{7}{l}{
			{\small LeNet-4 has only one FC layer; LeNet-1 has none.}}
	\end{tabular}}

%	\vspace*{-6mm}
\end{table} 

\vspace*{3mm}
\noindent\textbf{RQ5 (Layer Sensitivity).} 
Since \approach\ operates layer-wise, we investigated how IDC varies for different layers of a DL system. 
Table~\ref{tab:layer_sens} shows the coverage results for $m \in \{4, 6\}$ across layers, ordered by their depth for the three DL systems, using the original test set $U_O$ and that augmented with semantically diverse inputs $U_{O+DI}$. 
First, we observe that IDC value increases when the analysis is performed on deeper instead of shallow layers. 
For instance, in LeNet-4 and the $U_O$ test set, IDC increases from 33.9\% in Conv1 to 81.6\% in Conv2 until it reaches 86.0\% in FC1. 
We consider this observation as a confirmation of the ability of DL systems to extract more meaningful features in deeper layers. 

Furthermore, IDC is more sensitive to the test set with semantically diverse inputs ($U_{O+DI}$). 
In fact, we can observe a steady increase in the delta in IDC values between $U_{O+ID}$ and $U_O$ for more deeper layers. 
For Lenet-5, for instance, the IDC delta grows from 3.1\% in Conv1 to 5.6\% and 7.2\% in Conv2 and FC1, respectively, until it reaches 7.6\% for FC2. 
This behaviour persists despite the slight decrease in IDC value between FC1 and FC2 for both $U_O$ and $U_{O+DI}$. 
This observation reasserts our findings in RQ2 (cf. Table~\ref{tab:relevance_validation}).

\textbf{Overall, the chosen target layer affects the result of IDC. 
	Since the penultimate layer is responsible to understand se- mantically-important high-level features, we argue it is a suitable choice to assess the adequacy of a test set using IDC.}

\subsection{Threats to Validity}
We mitigate \textbf{construct validity} threats that could occur due to simplifications in the adopted experimental methodology using widely-studied datasets, i.e., MNIST~\cite{lecun1998mnist}, Cifar-10~\cite{cifar_model} and Udacity self-driving car challenge~\cite{udacity}. Also, we employed publicly-available  DL systems including LeNet~\cite{lecun1998mnist} and  Dave-2~\cite{bojarski2016end} that have different architectures and achieve competitive performance results~\cite{Goodfellow-et-al-2016}.
Also, we mitigate threats related to the identification of important neurons (Algorithm~\ref{alg:analysis}) by adapting techniques from the \textit{explainable AI} area for identifying input parts responsible for a decision~\cite{montavon2017explaining}.

We limit \textbf{internal validity} threats that could introduce bias when establishing the causality between our findings and the experimental study by designing independent research questions to evaluate \approach. 
Hence, we illustrate the performance of \approach\ in RQ1 and RQ2 for different values of important neurons $m \in \{6, 8, 10, 12\}$ and by augmenting the original test sets with both \textit{numerically diverse} and \textit{semantically diverse} perturbed inputs. 
The granularity of IDC increases exponentially with higher $m$ values, thus requiring a substantially larger number of inputs to be satisfied.
%In contrast to NC which can be satisfied with a test set containing the same number of test inputs as the number of neurons in the DL system~\cite{MJX18,Kim2019aa}. 
We also assessed the effectiveness of IDC to detect adversarial examples and confirmed its positive correlation with state-of-the-art coverage criteria for DL systems in RQ3 and RQ4, respectively. 
Furthermore, we investigate the effect of layer selection on IDC result in a structured manner in RQ5.
Finally, when randomness can play a factor (e.g., in RQ1 and RQ2), we reduce threats that the observations might have been obtained accidentally by reporting results over five independent runs per experiment.

%\textbf{Internal Validity:} The coverage measured by \approach mainly depends on three factors: subject layer, number of neurons considered as important and the number of clusters. For the first factor, we design a set of experiments (cf. RQ5) and investigate the effect of layer selection on the coverage ratio in a structured manner. For the second factor, we illustrate the performance of \approach\ in RQ1 and RQ2 for different number of important neurons. The number of inputs that is required to satisfy IDC can be large, especially, when the number of important neurons is higher. As DNNs implement very complex functions, a coverage criterion which requires large number of inputs to be satisfied would makes sense. For example, NC can be satisfied with a test set containing the same number of test inputs as the number of neurons in the DNN under test, in theory. On the other hand IDC may require millions of inputs which is more realistic when the complexity of the functionality (e.g. image recognition) is considered. Last, for the third factor, the number clusters is decided automatically by Silhouette which is an internal clustering validation index that computes the goodness of a clustering structure without external information.

We mitigate \textbf{external validity} threats that could affect the generalisation of IDC by developing \approach\ on top of the open-source frameworks Keras and Tensorflow which enable whitebox DNN analysis. 
We further reduce the risk that \approach\ might be difficult to use in practice by validating it against several DL systems trained on three popular datasets (MNIST~\cite{lecun1998mnist}, Cifar-10~\cite{cifar_model}, Udacity~\cite{udacity}). 
However, more experiments are needed to assess the performance of \approach\ using other techniques to identify the important neurons (e.g., DeepLift~\cite{shrikumar2017learning}), to extract clusters within important neurons (e.g., hierarchical clustering) and to validate the cohesion and separation of those clusters. 
These experiments are part of our future work.

%We address \textbf{external validity} threats that could affect the generalization of the \approach\ tool-supported instanceWe limit this threat by developing \approach\ using the open-source frameworks Keras and Tensorflow which enable whitebox DNN analysis. One can also examine other explainable AI techniques for IDC. We further reduce the risk that \approach\ might be difficult to use in practice by validating it against several DNN instances trained on three popular datasets. 

% !TEX root = ../main.tex

%\vspace*{-5mm}
\section{Related Work}
\label{sec:relatedWork}

% \noindent
% \textbf{DNN Testing and Verification.}

Trustworthiness issues in DL systems urged researchers to develop techniques that enable their effective and systematic testing~\cite{zhang2019machine}. 
Existing research in the area adapts testing techniques and criteria from traditional software engineering (e.g.,~\cite{MZSXJ18,SWRHKK18,eniser2019deepfault,ma2019deepct}) while other proposes novel test adequacy criteria~\cite{huang2018safety}. 
For instance, DeepXplore~\cite{pei2017deepxplore} introduces neuron coverage for measuring the ratio of neurons whose activation values are above a predefined threshold.
Similarly, DeepGauge~\cite{MJX18} introduces a family of adequacy criteria based on a more detailed analysis of neuron activation values.
DeepCT~\cite{ma2019deepct} proposes a combinatorial testing approach, while DeepCover~\cite{sun2018testing} adapts MC/DC from traditional software testing and defines adequacy criteria that investigate the changes of successive pairs of layers.
Recent research also proposes testing criteria and techniques driven by symbolic execution~\cite{GWZPK18}, coverage guided fuzzing~\cite{OG18,XMJCXLLZYS18} and metamorphic transformations~\cite{TPSR18}, while other research explores test prioritization~\cite{byun2019input} and fault localisation \cite{eniser2019deepfault}.% for DL systems. 

Although the objective of existing research is to guide testing of DL systems, eventually improving their accuracy and robustness, the majority concerns testing adequacy based on neuron-level properties. 
In contrast, \approach, driven by the fact that the behaviour of a DL system is determined layer-wise~\cite{Goodfellow-et-al-2016}, proposes a layer-wise and importance-based test adequacy criterion. 
In our experimental study (cf. Section~\ref{sec:evaluation}), we compare the performance of IDC against other layer-wise criteria (e.g., TKNC) and show that IDC is more informative.  
The recent research on using surprise adequacy to guide %DL system 
testing~\cite{Kim2019aa} is complementary to \approach.

A closely-related research branch is the provision of guarantees for the trustworthinesss of DL systems via formal verification~\cite{Huang18}. 
AI$^2$~\cite{GMDTCV18} uses abstract interpretation to verify safety properties, while~\cite{pulina2010abstraction} employs abstraction refinement. 
Other research uses SMT solvers to identify safe  regions in the input space and thus establish the robustness of DL systems~\cite{katz2017reluplex,gopinath2017deepsafe}.
Instead of SMT solvers, ReluVal~\cite{wang2018formal} finds bounds for security properties using interval arithmetic. %/ instead of SMT-solvers.
Finally, DLV~\cite{wicker2018feature} verifies local robustness based on user-defined manipulations.
\approach\ identifies important neurons using techniques from the \textit{explainable AI} area (e.g.,~\cite{bach2015pixel}); thus, it is  orthogonal to existing research on DL system verification.

%Formal DNN verification aims at providing guarantees for trustworthy DNN operation~\cite{Huang18}. Abstraction refinement is used in~\cite{pulina2010abstraction} to verify safety  properties of small neural networks with sigmoid activation functions, while  AI$^2$~\cite{GMDTCV18} employs abstract interpretation
%abstract domains and transformation operators 
%to verify similar properties. 

%Reluplex~\cite{katz2017reluplex} is an SMT-based approach that verifies safety  and robustness of DNNs with ReLU activation functions, and DeepSafe~\cite{gopinath2017deepsafe} uses Reluplex to identify safe  regions in the input space. DLV~\cite{wicker2018feature} can verify local DNN robustness given a set of user-defined manipulations. ReluVal~\cite{wang2018formal} formally checks security properties of DNNs  without using SMT solvers, instead they use interval arithmetic to give bounds on the DNN outputs.

Test adequacy is a widely-studied topic within traditional software engineering~\cite{pezze2008software}. 
Interested readers can find comprehensive reviews of relevant research in this area in related surveys and books \cite{grindal2005combination,jia2011analysis,ammann2016introduction,mathur2013foundations}. 
%Due to space limitation, we cannot provide a comprehensive review of relevant research. Instead, we refer interested readers to 

% \noindent
% \textbf{Test Adequacy Criteria in Traditional Software.}
%Test adequacy in the field of software engineering is a well studied topic and there are substantial amount of works in the literature. Therefore, we can not provide all of them here. Here, we note several related surveys and books \cite{grindal2005combination,jia2011analysis,ammann2016introduction,mathur2013foundations}, so that an interested reader can benefit from them.
% !TEX root = ../main.tex

%\vspace*{-5mm}
\section{Conclusion}
\label{sec:conclusion}

Ensuring the trustworthiness of DL systems 
%by reducing the residual risk for unexpected behaviour 
requires their thorough and systematic testing.  
%We contribute in this effort, by introducing 
\approach\ is a systematic testing methodology reinforced by an \textit{Importance-Driven} (IDC) test adequacy criterion for DL systems.
\approach\ analyses the internal neuron behaviour to create a \textit{layer-wise functional understanding} 
%based on the most important neurons and a finite set of
and automatically establish a finite set of clusters that represent the behaviour of the most important neurons to an adequate level of granularity. 
The {Importance-Driven} adequacy criterion  measures the adequacy of a test set as the ratio of combinations of important neurons clusters covered by the set.
Our experimental evaluation shows that IDC achieves higher results for test sets with \textit{semantically-diverse} inputs. 
IDC is also sensitive to adversarial inputs and, thus, effective in detecting misbehaviour in test sets. 

Our future work involves (1) investigating methods to automatically determine the number of important neurons; (2) improving the robustness of IDC; (3) evaluating \approach\ on other DL systems and datasets; and (4)  examining how \approach\ results can be incorporated into safety cases~\cite{calinescu2018engineering,kelly1999arguing}.

\section*{Acknowledgements}
This research was supported in part by Semiconductor Research Corporation under task 2020-AH-2970.

\bibliographystyle{ACM-Reference-Format}
\bibliography{main}

\end{document}